# The sounds of science – a symphony for many instruments and voices

**Gerianne Alexander**[1], **Roland E. Allen**[2], **Anthony Atala**[3], **Warwick P. Bowen**[4,5], **Alan A. Coley**[6], **John Goodenough**[7], **Mikhail I. Katsnelson**[8], **Eugene V. Koonin**[9], **Mario Krenn**[10,11], **Lars S. Madsen**[5], **Martin Månsson**[12], **Nicolas P. Mauranyapin**[4], **Art I. Melvin**[10,13], **Ernst Rasel**[14], **Linda E. Reichl**[15], **Roman Yampolskiy**[16], **Philip B. Yasskin**[17], **Anton Zeilinger**[13,10] and **Suzy Lidström**[18]

[1]Department of Psychological and Brain Sciences, Texas A&M University, College Station, TX, USA

E-mail: `galexander@tamu.edu` (Gerianne Alexander)
[2]Department of Physics and Astronomy, Texas A&M University, College Station, Texas, U.S.A.

E-mail: `allen@physics.tamu.edu` (Roland E. Allen)
[3]Wake Forest Institute for Regenerative Medicine, 391 Technology Way, Winston-Salem, 27157, NC, USA

E-mail: `regenmed@wakehealth.edu` (Anthony Atala)
[4]School of Mathematics and Physics, University of Queensland, St. Lucia 4072, Australia

E-mail: `n.mauranyapin@uq.edu.au` (Nicolas P. Mauranyapin)
[5]Australian Centre for Engineered Quantum Systems, University of Queensland, St. Lucia 4072, Australia

E-mail: `w.bowen@uq.edu.au` (Warwick P. Bowen)

E-mail: `m.lars@uq.edu.au` (Lars S. Madsen)
[6]Department of Mathematics and Statistics, Dalhousie University, Halifax, B3H 4R2, Nova Scotia, Canada

E-mail: `aac@mathstat.dal.ca` (Alan A. Coley)
[7]Walker Department of Mechanical Engineering, Cockrell Institute, University of Texas at Austin, Austin, Texas

E-mail: `jgoodenough@mail.utexas.edu` (John Goodenough)
[8]Institute for Molecules and Materials, Radboud University, Nijmegen, 6525AJ, The Netherlands
[9]National Library of Medicine, National Center for Biotechnology Information, Bethesda, MD 20894, USA
[10]Institute for Quantum Optics and Quantum Information (IQOQI), Austrian Academy of Sciences, Boltzmanngasse 3, Vienna, 1090, Austria

E-mail: `anton.zeilinger@univie.ac.at` (Anton Zeilinger)
[11]Department of Chemistry, University of Toronto, Toronto, Canada

E-mail: `mario.krenn@univie.ac.at (Mario Krenn)`
[12]Dept. of Applied Physics, KTH Royal Institute of Technology, Stockholm, SE-164 40 Kista, Sweden

E-mail: `condmat@kth.se (Martin Månsson)`
[13]Vienna Center for Quantum Science & Technology (VCQ), Faculty of Physics,, University of Vienna, Boltzmanngasse 5, Vienna, 1090, Austria

E-mail: `art.i.melvin@gmail.com (Art I. Melvin)`
[14]Institut für Quantenoptik, Welfengarten 1, Hannover, 30167, Germany

E-mail: `rasel@iqo.uni-hannover.de (Ernst Rasel)`
[15]Center for Complex Quantum Systems and Department of Physics, The University of Texas at Austin, Austin, Texas, U.S.A.
[16]Department of Computer Engineering and Computer Science, Duthie Center for Engineering, University of Louisville, Louisville, 40292, Kentucky, USA

E-mail: `roman.yampolskiy@louisville.edu (Roman Yampolskiy)`
[17]Department of Mathematics, Texas A&M University, College Station, Texas, U.S.A.

E-mail: `yasskin@math.tamu.edu (Philip B. Yasskin)`
[18]Department of Physics and Astronomy (Visiting scholar at time of writing), Texas A&M University, College Station, Texas, USA

E-mail: `suzy.lidstrom@gmail.com (Suzy Lidström)`

**Abstract.** *Sounds of Science* is the first movement of a symphony for many (scientific) instruments and voices, united in celebration of the frontiers of science. John Goodenough, the maestro who transformed energy usage and technology through the invention of the lithium-ion battery, opens the programme, reflecting on the ultimate limits of battery technology. This applied theme continues through the subsequent pieces on energy related topics – the sodium-ion battery and artificial fuels, by Martin Månsson – and the ultimate challenge for 3D printing, the eventual production of life, by Anthony Atala. A passage by Gerianne Alexander follows, reflecting on a related issue: How might an artificially produced human being behave? Next comes a consideration of consiousness and free will by Roland Allen and Suzy Lidström. Further voices and new instruments enter as Warwick Bowen, Nicolas Mauranyapin and Lars Madsen discuss whether dynamical processes of single molecules might be observed in their native state.

The exploitation of chaos in science and technology, applications of Bose-Einstein condensates and a consideration of the significance of entropy follow in pieces by Linda Reichl, Ernst Rasel and Roland Allen, respectively. Mikhail Katsnelson and Eugene Koonin then discuss the potential generalisation of thermodynamic concepts in the context of biological evolution.

Entering with the music of the cosmos, Philip Yasskin discusses whether we might be able to observe torsion in the geometry of the universe. The crescendo comes with the crisis of singularities, their nature and whether they can be resolved through quantum effects, in the composition of Alan Coley. The climax is Mario Krenn, Art Melvin and Anton Zeilinger's consideration of how computer code can be autonomously surprising and creative. In a harmonious counterpoint – his 'Guidelines for considering AIs as coauthors' – Roman Yampolskiy concludes that such code is not yet able to take responsibility for coauthoring a paper. An interlude summarises a speech by Zdeněk Pa-poušekof.

In a subsequent movement, new themes emerge as we seek to comprehend how far we have travelled along the path to understanding, and speculate on where new physics might arise, by glancing at what the history of science has to teach us: Who would have imagined, 100 years ago, a global society permeated by smartphones and scientific instruments so sophisticated that genes can be modified and gravitational waves detected?

## 1. Prelude by Suzy Lidström

In the title song of *Sounds of Silence,* Paul Simon confronts his audience with a complete breakdown in communication in a society where, in his words, people talk without speaking and hear without listening, and where songs are never shared. It is our hope that, at a time when scientific papers have become increasingly specialised, are considerably more numerous and, consequently, are less well read, the voices united in this piece will be heard, shared, and enjoyed as much as the grand challenges facing the scientific community have been in our previous publications and those of other authors ([1], [2], [3], [4], [5]).

In this score, many voices are joined in an exploration of the stimulating themes proposed for consideration by the participants of a conference held in the Czech Republic, in Prague in 2017. At that meeting, members of the scientific community were invited to share the challenges they perceived to be most pressing and would like to see resolved in their lifetimes, or contribute questions that they found fascinating. Through the resultant questions and the willingness of experts to respond to them, we find evidence of a common concern in the scientific community for the issues of our time – a stark contrast to the self-induced isolation of the citizens who dared not disturb the *Sounds of Silence.* Insights are gained as each expert explains the enquiry raised, clarifies the issues for those unfamiliar with them, and presents the particular challenges in relation to his or her own area of expertise. As the authors reflect on the progress made towards the eventual resolution of the issues, we not only appreciate how far the fields have advanced, but see how the research community can contribute to furthering our understanding in the future. We hope that our audience will enjoy a front row seat at a rare performance in a specialised world.

It is fitting that the title of this manuscript reflects a musical theme for diverse reasons, not least of which is the fact that physicists and mathematicians often harbour a deep-seated love of music. In this opening section, this is exemplified by (the albeit exceptional) Albert Einstein, who passed seventeen fruitful months in Prague, where he: "... found the necessary composure" to develop the basic ideas underpinning the theory of general relativity [6]. We seek to emulate the successful harmonisation of science and music that Einstein achieved in this city through the present set of forward-looking compositions: Each was written by an expert in response to a question stimulated by those posed by the participants attending *Frontiers in Quantum and Mesoscopic*

*Thermodynamics,* a conference with a truly brilliant evening programme of world-class musical performances worthy of the musical legacy of this city.

Einstein, the father of relativity, a major player in the development of quantum physics and a keen amateur musician, contemplated several of the themes that recur in their modern form in the movements of *Sounds of Science.* During the most productive period of his life, Einstein could be found enjoying the sounds of music in the salon Bertha Fanta. Throughout his sojourn in Prague Einstein took pleasure in making music in the company of the Winternitz family, and in particular, playing alongside piano teacher Ottilie Nagel, a sister-in-law of Professor Winternitz. Einstein's instrument of choice then and throughout his life was 'die violine', or *Lina* for short, the name he adopted for a succession of his violins Figure 1.

Einstein claimed to have spent some of the most beautiful moments of his life in Prague, notably in association with a visit by his friend and colleague, Paul Ehrenfest Figure 2.

In addition to developing a core of physics-related content, the issues raised by those who contributed the questions shown in Figure 3 stretch beyond the frontiers of quantum and mesoscopic physics. The grand challenges relating to climate change and our planet have, therefore, been put aside for a future publication, but here two authors address one critical component – the battery technologies that are required for the success of clean energy from sun and wind. We include tantalising topics in applied domains other than energy storage, such as bioprinting and eventual uses of Bose-Einstein condensates and chaos. On the theoretical side, questions examining some of the great mysteries in physics and cosmology are accompanied by careful explanations with relevant references, hopefully ensuring that even the most esoteric topics become accessible.

John Goodenough Figure 4, principal inventor of the lithium-ion battery, opens the programme by considering how far technology can be pushed, responding to the question: *What is the ultimate limit to battery technology?* This battery-related theme reverberates in Martin Månsson's recommendation for the widespread adoption of the sodium-ion battery in *Will lithium become the new oil?* and his accompanying piece, a consideration of the desirability of developing complementary technology: *Towards safe sodium batteries.*

Anthony Atala's composition *Will 3D printing be used to produce life?* contemplates this seemingly incredible question relating to an eventuality where life might be created on demand. Atala seizes on this alluring theme and makes it his own, clarifying how far we have progressed towards producing life and the manner in which we might be able to achieve it. Gerianne Alexander further enhances this theme, elaborating on the behavioural issues an artificial life-form could present. In *If a human were created atom by atom, molecule by molecule, would it behave like you and me?*, she elaborates on those people hearing without listening, talking without speaking to whom Paul Simon referred.

The Czech chemist Antonin Holy, inventor of the retroviral drugs that have been so successful for treating HIV, commented on the driving force behind our search for

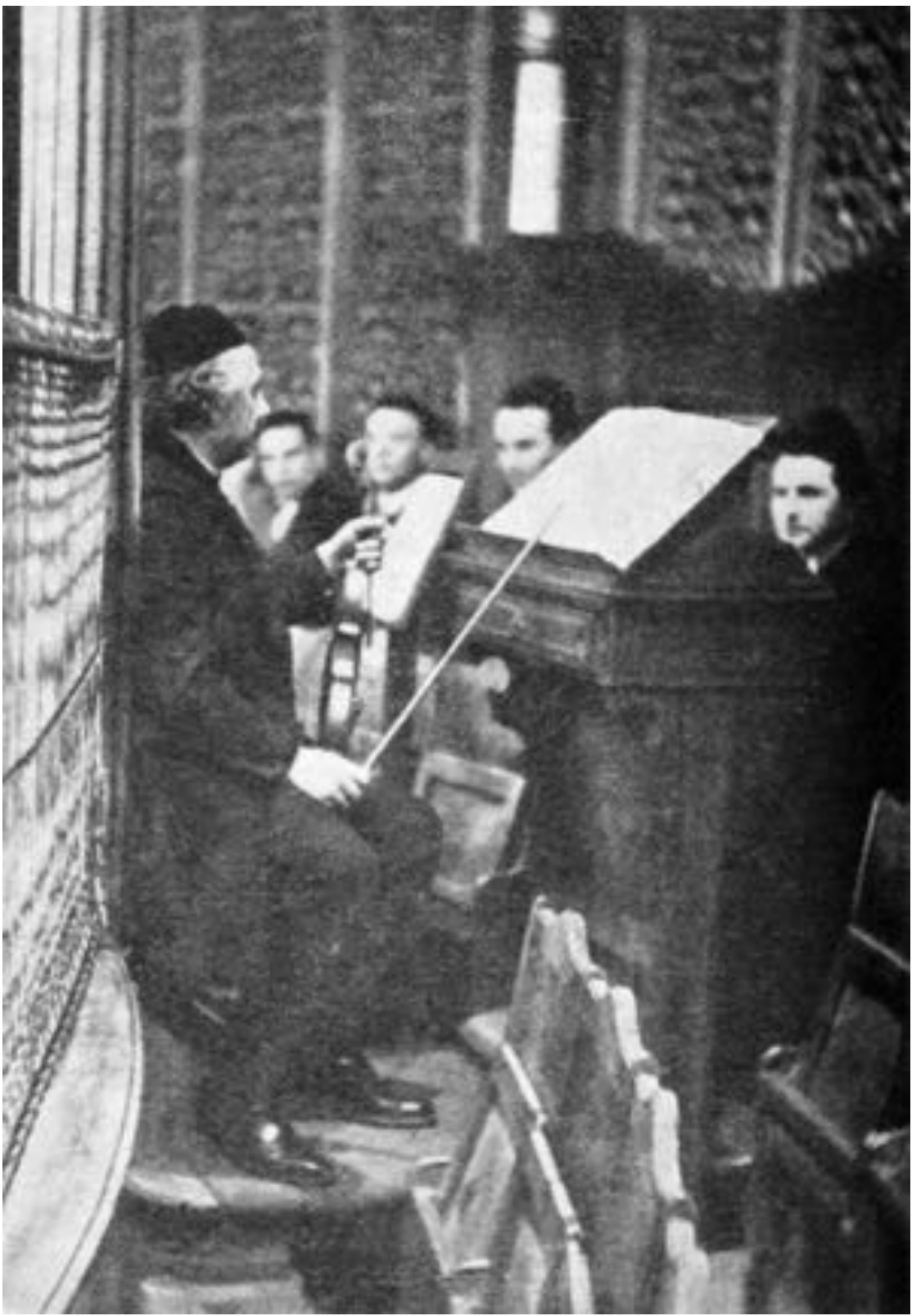

**Figure 1.** Albert Einstein playing his violin, *Lina*, at a charity concert in the New Synagogue in Berlin on $29^{th}$ January, 1930. Photographer: Anonymous [Public domain], via Wikimedia Commons.

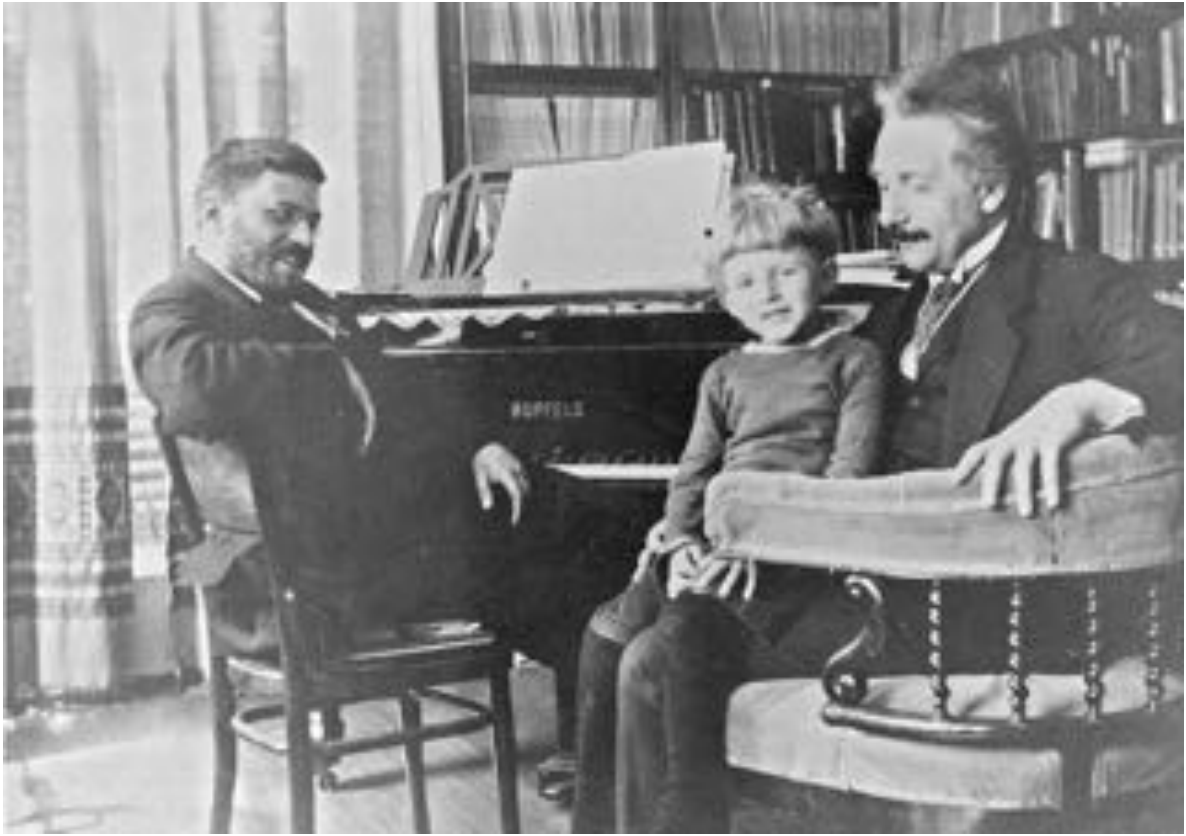

**Figure 2.** Einstein with Paul Ehrenfest. In 1929, Einstein was quoted as saying: "If I were not a physicist, I would probably be a musician. I often think in music. I live my daydreams in music. I see my life in terms of music... I get most joy in life out of my violin." [7] Many scientists acknowledge a similar depth of feeling for music. Victor Weisskopf, for instance, wrote, "Science became my profession, but music remains my religion." [8] Photographer: Anonymous [Public domain], via Wikimedia Commons.

increased understanding. It was, he said:

> The desire for knowledge and for overcoming the ordinary; a creative approach, intuition, enthusiasm, commitment and sacrifice that always has been, is and will be the driving force of human cognition.
>
> *Antonin Holy, Research and Development in the Czech Republic*

That informs us *of*, but does not explain, one of the greatest mysteries of the present era, human cognition. *What is consciousness and do we have free will?* forms the subject of a contribution by Roland Allen and Suzy Lidström.

Warwick Bowen, Nicolas Mauranyapin and Lars Madsen's return to molecular considerations, as they delve into the dominant theme of the conference, contemplating how we might attempt to hone in on the detailed dynamics of as yet inaccessible realms in: *Can quantum techniques tell us about the dynamics of single molecules in their native state?*

Successive pieces extend and elaborate on pure physics: Linda Reichl's composition emphasises the thermodynamic theme of the conference as she reflects on the question: *How can chaos be exploited in science and technology*? Ernst Rasel considers eventual future applications in *Do Bose-Einstein condensates of cold quantum gases have any practical applications?*, presenting quantum gravimetry and inertial sensing with enviable concision. Roland Allen then asks: *What does entropy mean and why is it so important?*, reminding us that, despite its central role throughout science and technology, entropy does not appear in the most fundamental laws of nature.

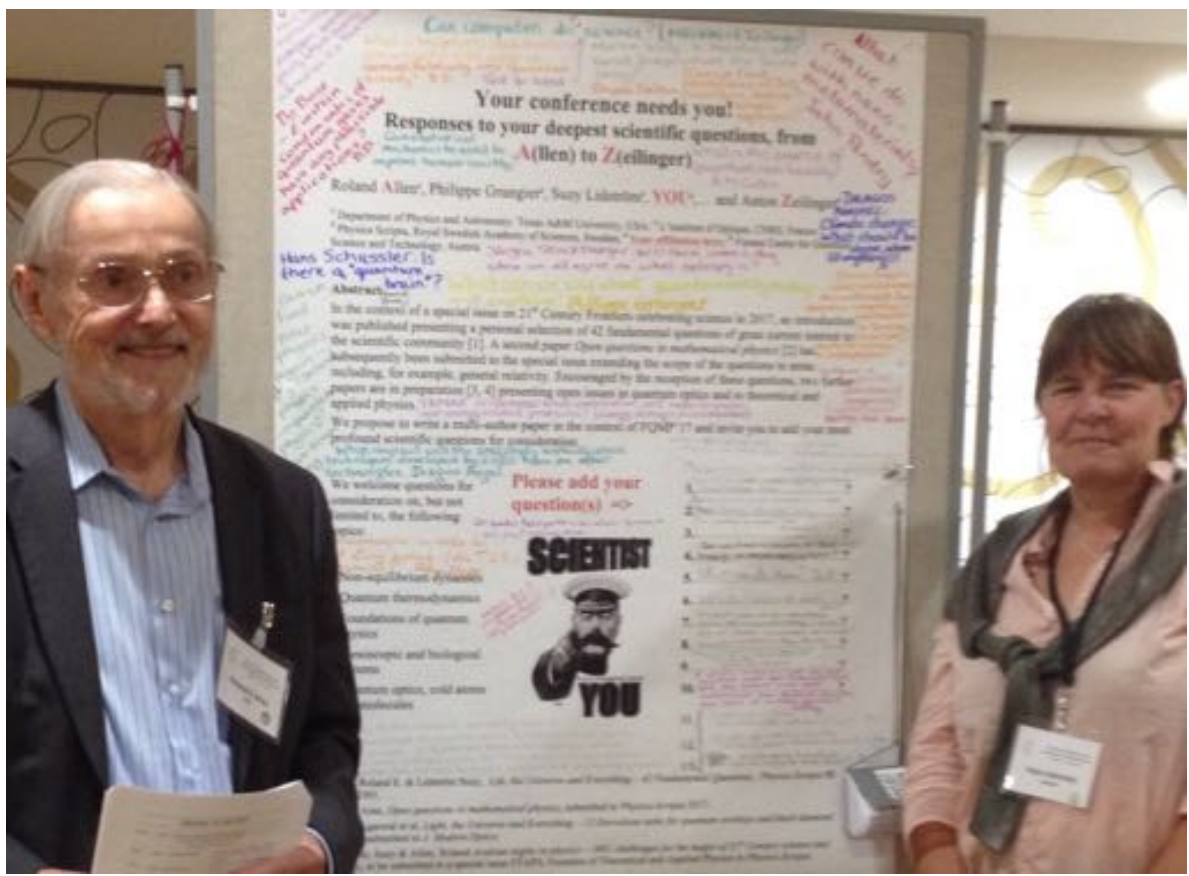

**Figure 3.** Two of the authors of this paper invited participants of Frontiers of Quantum and Mesoscopic Thermodynamics (FQMT) to consider the greatest challenges facing the scientific community at this time. All willing contributors were encouraged to propose questions for discussion allied with the frontiers of quantum and mesoscopic thermodynamics or to clarify the questions that they would most like to see resolved within their respective lifetimes. The themes developed here are responses to some of the questions posed. Credit: Suzy Lidström.

Phil Yasskin takes centre stage with: *Can we observe the torsion of the connection in the geometry of the universe?*, in a treatment of this advanced topic for the non-specialist.

The diversity of the repertoire becomes fully apparent as a touch of digital hardcore accompanies the voices of Mario Krenn and Anton Zeilinger. Together, they respond to the question: *How can a computer find autonomously new, surprising or creative* *solutions or insights?* Echoing this theme, Roman Yampolskiy considers *Guidelines for including AIs (forms of Artificial Intelligence) as co-authors* and we are reluctantly forced to conclude that computer code is not yet capable of taking responsibility for this paper.

Recapitulation: This rich symphony of ideas reflects the culture of contemporary science, the dramatic opposite of the impoverished society portrayed in Simon and Garfunkle's "Sounds of Silence".

We invite you to enjoy a performance in which we have attempted to emulate the spirit of discovery encapsulated by Victor Weisskopf when he said: "The joy of insight is a sense of involvement and awe, the elated state of mind that you achieve when you have grasped some essential point; it is akin to what you feel on top of a mountain after a hard climb or when you hear a great work of music."

## 2. What are the ultimate limits to battery technology? by John Goodenough

The changes that have taken place in battery technology over the last 60 years teach us always to keep an open mind for surprises. Today, battery technology is about to be transformed again by the advent of a dielectric amorphous-solid electrolyte.

A battery and an electrochemical capacitor consist of one or more identical electrochemical cells. On discharge, each cell delivers electric power $P_{dis} = I_{dis}V_{dis}$ for a time $\Delta t$. The total charge delivered is the cell capacity $Q(I_{dis})$ per unit weight or volume of the cell: at a constant $I_{dis} = \frac{dq}{dt}$

$$Q(I_{dis}) = \int_0^{\Delta t} (I_{dis}) \, \mathrm{d}\, t = \int_0^{QI_{dis}} \mathrm{d}\, q \tag{1}$$

and the density of stored energy is

$$\Delta E = \int_0^{\Delta t} P_{dis} \, \mathrm{d}\, t = \langle V_{dis} \rangle Q(I_{dis}) \tag{2}$$

The cells are connected in series to deliver a desired $V_{dis}$ and in parallel for a desired $I_{dis}$. An electrochemical capacitor stores electric power as electrostatic energy and a conventional battery stores electric power as chemical energy; the cells of a hybrid battery store both chemical and electrostatic energy.

The components of an electrochemical battery cell are two electronically conducting electrodes, an anode and a cathode, separated by an electrolyte. Both the chemical (faradaic) and electrostatic (capacitive) components of stored energy in a battery have, on charge and discharge, both an ionic and an electronic component; the electrolyte conducts the ionic component inside the cell, but it is an electronic insulator to force the electronic component to traverse an external circuit to give an electronic current $I$ between the two electrodes at a voltage $V$ for a time $\Delta t$. The ionic component in the electrolyte may be a mobile cation current or a displacement current associated with electric dipoles. If the electrolyte is a liquid, the electrodes are kept apart by a separator that is neither reduced by a reductant anode nor oxidized by an oxidant cathode; if the electrolyte is a solid, it is also the separator.

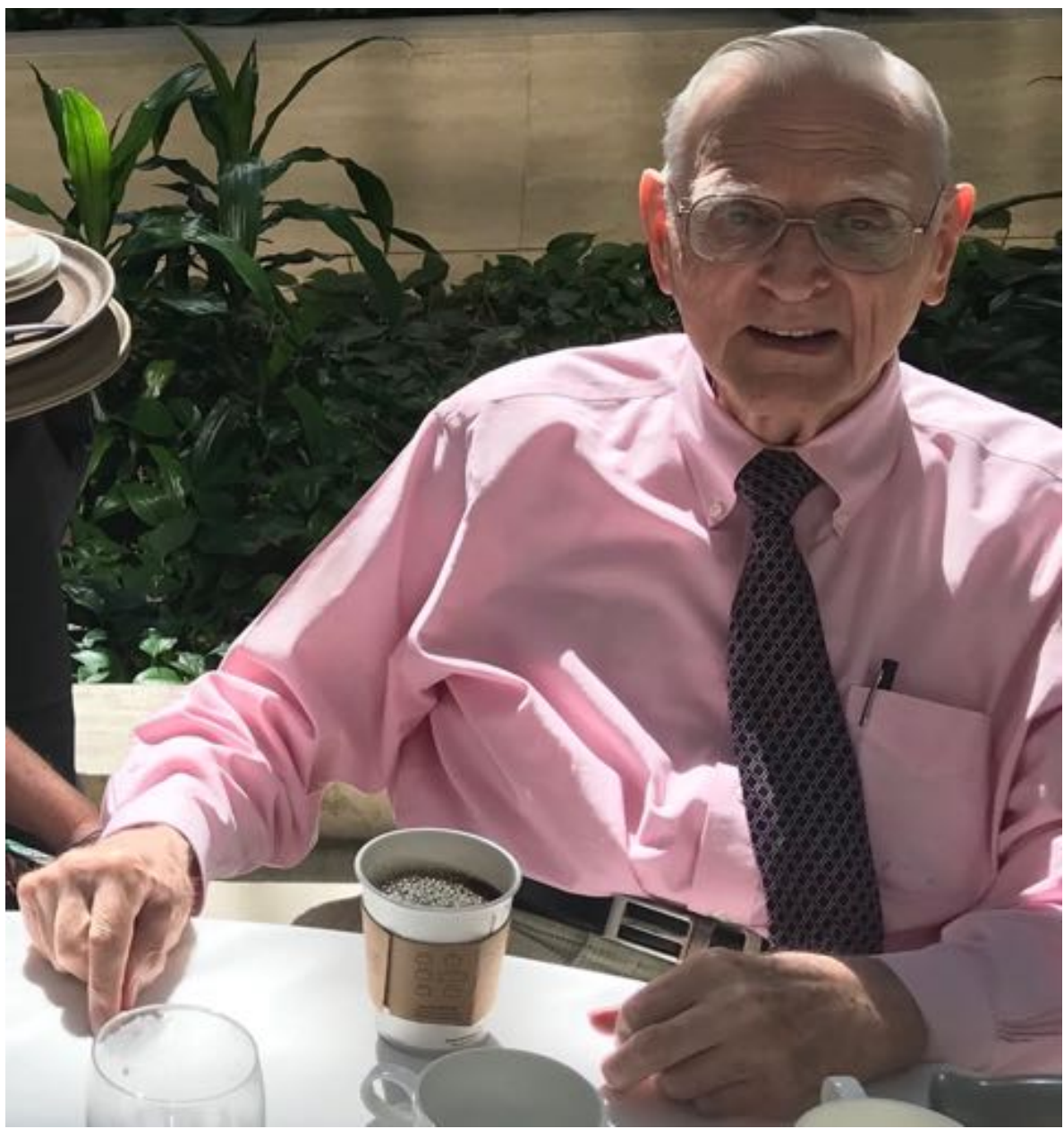

**Figure 4.** John Goodenough, principal inventor of the lithium-ion battery on the occasion of the delivery of his acceptance speech for the Robert A. Welch Award for his contributions to chemistry and mankind. Almost all portable modern devices include a lithium-ion battery. Photograph: Roland Allen.

Implicit in the question of an "ultimate limit" to battery technology is the limit to a rechargeable battery. The faradaic component of a rechargeable cell requires a reversible chemical reaction between the two electrodes. The chemical reaction of an electrode involves either reversible plating/stripping from/to the electrolyte or the reversible insertion/extraction of the working cation into a solid or a molecule. If the molecule is oxygen gas at an air cathode, the electrode must support catalysts for the oxygen-reduction reaction (ORR) and the oxygen-evolution reaction (OER). The capacity $Q(I_{dis})$ of an insertion-reaction electrode is limited by the reversible solid-solution range of the electrolyte mobile (working) cation in the electrode, and this range depends on the rate of insertion and, therefore, on $I_{dis}$.

The efficiency of electric-power is $P_{dis}/P_{ch} < 100\%$ since $V_{dis} = V_{oc} - \eta_{dis}I_{dis}$ and $V_{ch} = V_{oc} + \eta_{ch}I_{dis}$; $V_{oc} = (\mu_A - \mu_c)/e$ is the open-circuit ($oc$) voltage, $e$ is the magnitude of the electronic charge, and $(\mu_A - \mu_c)$ is the difference between the chemical energies (Fermi levels) of the anode and the cathode. The coulomb ($cm$) efficiency of a charge/discharge cycle taken at a fixed control current $I_{cm}$ is $Q(I_{con})_{dis}/Q(I_{con})_{ch}$. It has recently been shown that this efficiency can be greater than 100% with a dielectric solid electrolyte.

The factors that limit battery technology are the *cost* of multicell batteries and the *energy density* of a fast *charge/discharge*. The electrolyte is the principal component that controls these factors; and herein a history of the evolution of the cell electrolyte follows a review of why the electrolyte has been the critical component.

The ionic conductivity in the cell electrolyte is orders of magnitude smaller than that of the electrons in a metallic external circuit. The diffusion of the ions in the electrolyte increases with $\sigma_M\ A/d$, where $\sigma_M$ is the conductivity of the mobile cation in the electrolyte and $A/\mathrm{d}$ is the ratio of area to thickness of the cell. Therefore, a $\sigma_M > 10^{-3}$ S cm$^{-1}$ and the ability to create a large-area electrolyte with a thickness $d \lesssim 30$ μm is required for an acceptable charge/discharge rate.

The number of charge/discharge cycles before the capacity of a cell decreases to 80% of its initial value represents the *cycle life* of a cell; and a long cycle life (tens of thousands of cycles) is needed for a cost that can compete with the energy stored in a fossil fuel. An electrolyte has an energy gap $E_g$ between its lowest unoccupied molecular orbital (LUMO or bottom of conduction band) and its highest occupied molecular orbital (HOMO or top of valence band). If the anode Fermi level is $\mu_A > $ LUMO, the electrolyte is reduced and if the cathode Fermi level is $\mu_C < $ HOMO, the electrolyte is oxidized; unless a solid-electrolyte interphase (SEI) passivates the electrode/electrolyte reaction; the SEI must conduct the working ion of the cell. During charge, the voltage of a cell may increase $\mu_A$ or decrease $\mu_C$ to where the electrolyte is reduced or oxidized. Where $\mu_A$ or $\mu_C$ are at energies outside the electrolyte $E_g$, the formation of an SEI that changes its area on cycling reduces the cycle life of a cell.

Where the chemical reaction in a cell is between two solid electrodes, the electrodes change volume during charge/discharge. Where the faradaic reaction at an electrode/electrolyte interface consists of reversible plating/stripping of the electrode and the electrode wets a solid electrolyte or solid component of a composite electrolyte, the volume change of the electrode is constrained by strong electrode-electrolyte bonding to confine the volume change to perpendicular to the electrode/electrolyte interface; this volume change can be accommodated by a cell design that contains a spring that applies pressure perpendicular to the large-area interface. However, the volume change of a solid electrode particle into which the working cation is inserted/extracted reversibly, whether an alloy or an insertion compound, is three-dimensional. In this case, the electrolyte that contacts the electrode must be plastic enough to retain good contact on cycling; the electrolyte contacting the electrode must be plastic enough to accommodate the volume change, as occurs with a liquid or plastic-polymer electrolyte.

The electrode/electrolyte interface is also a heterojunction across which the Fermi level of two materials in contact is equalized by the formation of an electric-double-layer capacitor (EDLC). At open-circuit, the external electronic current is stopped, but the mobile (working) cation of the electrolyte can move to the anode interface to create an EDLC at the two electrolyte/electrode heterojunctions, which makes the anode the negative terminal and the cathode the positive terminal.

In 1960, batteries were being fabricated with a strongly acidic or alkaline aqueous electrolyte having a fast $H^+$ conductivity. However, water is separated into $H_2$ and $O_2$ on the application of a voltage $V \gtrsim 1.23$ V. With an $E_g$ of 1.23 eV, a stable rechargeable battery with an aqueous electrolyte has a $V_{dis} \lesssim 1.5$ V. Although a rechargeable lead-acid battery cell ($PbO_2/H_2SO_4/Pb$) has a $V_{dis} = 2$ V, it self-discharges over time with the precipitation of $PbSO_4$. The best cell in 1960 had a layered $NiOOH$ charged cathode with an alkaline (KOH) electrolyte and a Cd anode. The $Ni^{3+}/Ni^{2+}$ redox energy of the $NiOOH + H^+ + e^- = Ni(OH)2$ reaction is well-matched to the HOMO of a KOH solution, and Cd has a Fermi level well-matched to the LUMO in this nickel-cadmium battery cell. However, with a $V_{dis} < 1.5$ V, a larger energy density requires, according to Eq. (2), a larger $Q(I_{dis})$. The limiting rechargeable battery with an aqueous electrolyte would be an air/zinc battery containing low-cost ORR and OER electrocatalysts on a chemically stable metallic support. This cell is possible, but an air electrode is not a feasible option for an electric road vehicle.

During the 1960s, Jean Rouxel in France and Robert Schöllhorn in Germany were exploring the chemistry of reversible $Li^+$ intercalation into layered sulfides $MS_2$ (M is a transition metal). In 1967, Kummer and Webber discovered good $Na^+$ conductivity at 300 °C in a solid ceramic and invented the sodium sulfur rechargeable cell that used molten sodium as the anode, carbon felt in molten sulfur as the cathode, and their ceramic $Na^+$ conductor as the electrolyte; it operated at 350 °C and turned out to be too expensive to maintain. However, this development and the oil crisis of the early 1970s stimulated thinking about rechargeable batteries with a different electrolyte.

A primary (non-rechargeable) cell with an organic liquid-carbonate $Li^+$ electrolyte and a lithium anode pacified by an SEI had been marketed, so it was suggested at a conference by Brian Steele of England that a TiSq cathode with a metallic lithium anode might make a rechargeable battery with a higher energy density. In 1976, M. Stanley Whittingham demonstrated that a $TiS_2/Li$ coin cell gave a $V_{dis} \approx 2.2$ V with an acceptable rate of charge/discharge. The Exxon-Mobile Corporation licensed the concept and hired Whittingham to develop a marketable cell. However, cell fires, even explosions, soon shut the effort down. During charge in a rechargeable cell, plating a metallic anode develops anode dendrites that, on repeated charges, grow across a thin electrolyte to the cathode to create an internal short-circuit and thermal runaway with incendiary consequences.

The solution to this problem was to fabricate a discharged oxide as cathode and to investigate how much $Li^+$ can be extracted from the layered $LiCoO_2$ or $LiNiO_2$. These oxides gave a $V_{dis} \simeq 4.0$ V versus lithium, which turned out to be well-matched to the HOMO of the flammable liquid-carbonate electrolyte. This voltage would allow the development of a discharged anode; but without identification of the discharged anode, battery companies would not take the risk of licensing the concept. However, to avoid dendrites from the anode, chemists were studying the reversible intercalation of $Li^+$ into graphitic carbon. In Japan, Akira Yoshino recognized that graphitic carbon offered a discharged anode into which $Li^+$ can be intercalated dendrite-free, and the SONY Corporation licensed the Li-ion battery with a $LiMO_2$ (where M is Co or Ni) cathode and a graphitic-carbon anode to power the first cell telephone, thereby launching the wireless revolution.

Although the Li-ion battery with a carbon anode and a $LiCoO_2$ cathode has enjoyed a great financial success and has been used to demonstrate that an all-electric road vehicle powered by a rechargeable battery can provide a vehicle performance competitive with that of a vehicle powered by an internal combustion engine, its cost and the flammable liquid electrolyte has prevented achieving the goal of a safe, low-cost power source with a fast charge and a sufficient volumetric energy density.

The solution would appear to be a solid electrolyte with a large enough $E_g$ that it is not reduced by a lithium anode and is not oxidized at a charging voltage $V_{ch} \gtrsim 5$ V. The logical solid electrolyte would appear to be a polymer; a thin large-area ceramic electrolyte would be too brittle. Although mechanically robust, polymer-ceramic-composite $Li^+$ and $Na^+$ electrolytes with a $\sigma_M \lesssim 5$ x $10^{-3}$ S cm$^{-1}$ have now been made and it has been shown that plating/stripping reversibly of dendrite-free alkali-metal anodes can be achieved with the composite electrolyte, the best cells to date operate at a temperature $T_{op} > 60$ °C because of a $\sigma_M \lesssim 5$ x $10^{-3}$ S cm$^{-1}$ and a d $>$ 30 μm. A solution to this problem shows promise; it consists of a dielectric amorphous ceramic (glass) developed by M. Helena Braga of the University of Porto, Portugal. The Braga glass has a room-temperature $Li^+$ or $Na^+$ ionic conductivity of $\sigma_M > 10^{-2}$ S cm$^{-1}$ which is comparable to that of the flammable liquid electrolyte and it retains a $\sigma_M > 10^{-3}$ S cm$^{-1}$ down to $-30$ °C. As a composite with a polymer, it can be made mechanically robust by mixing a polymer with the glass particles. Moreover, reversible plating/stripping at a negligible impedance for thousands of cycles with no capacity fade has been demonstrated with symmetric cells. A dendrite-free alkali-metal anode provides the limiting anode capacity of a rechargeable cell. The ability to plate/strip an alkali-metal electrode suggested that an asymmetric cell could be made in which an alkali-metal anode is transferred reversibly between the anode and the cathode at a finite voltage, and a $V_{dis} \simeq 3$ V has been demonstrated. This electrolyte would appear to provide an ultimate solution, but what capacity can be achieved and at what rate of plating/stripping has yet to be determined. This solution need not be confined to the Braga glass if another dielectric solid electrolyte with a $\sigma_M > 10^{-2}$ S cm$^{-1}$ at 25 °C is found.

The coexistence of a fast-moving working $Li^+$ or $Na^+$ cation and slower-moving electric dipoles in the dielectric solid electrolyte provides the novel phenomena of self-charge and self-cycling. The self-cycling is the first example of an electrochemical relaxation oscillator. These phenomena have allowed the demonstration of a high-voltage cell that combines both a fast capacitive component and a large-capacity faradaic component. A coin cell with an insertion-compound cathode has been cycled rapidly for over 10,000 times with a coulomb efficiency in excess of 100%.

In summary, it is still premature to define the "ultimate limits" to rechargeable battery technology.

## 3. Will Lithium be the New Oil? - Towards Green & Safe Sodium Batteries by Martin Månsson

### 3.1. Lithium-Ion Batteries (Li)

In our modern society one of the main scientific and technical challenges is finding out how to convert and store clean energy. For portable applications, i.e. electrical automobiles, smart-phones, tablets etc., rechargeable lithium-ion (Li-ion) batteries (LIBs, discussed by John Goodenough in the previous section) are the backbone of current technologies. LIBs are electrochemical cells that directly convert chemical energy into electrical energy ('electricity'). One of the main reasons for the general success of electrochemical devices (batteries, fuel-cells etc.) is that such conversion is extremely efficient (up to 98%).

A significant obstacle for electric cars to reach maturity has long been the development of a sufficiently high-capacity, cheap, lightweight and safe rechargeable battery. However, LIBs have gone through a dramatic improvement during the last few decades, and today electric automobiles are starting to break through on the open market. At the same time, demand for the raw materials needed to produce the batteries has come to be emphasized as a significant issue. Most often, discussions have been centered around the toxic Cobalt (Co) that is often extracted by children in the mines of *e.g.* Congo [9], [10]. However, the other main element of LIBs, the lithium itself, has also started to come into focus. Lithium is a rather rare metal, existing in a mere 17 ppm concentration in the earth's crust (see, e.g., [11]). In addition, known lithium deposits are very unevenly distributed [12] between the different continents (see Figure 5 (a)). With a strongly increasing demand for raw material the price of lithium has more than tripled [13] and [14] during the last couple of years (see Figure 5 (b)). Furthermore, the environmental impact from lithium extraction and difficulties around recycling is highly debated. Lithium deposits are usually found in salt flats where water supply tends to be limited, however the mining process itself requires substantial amounts of water as well as a series of chemicals for leaching purposes. As a result, contamination and depletion of scarce water supplies can lead to substantial impact on the local environment as well as on the population. Finally, the transport of raw materials and production of the batteries actually causes a significant $CO_2$ footprint for the electrical vehicles, as recently presented in a Swedish report [15].

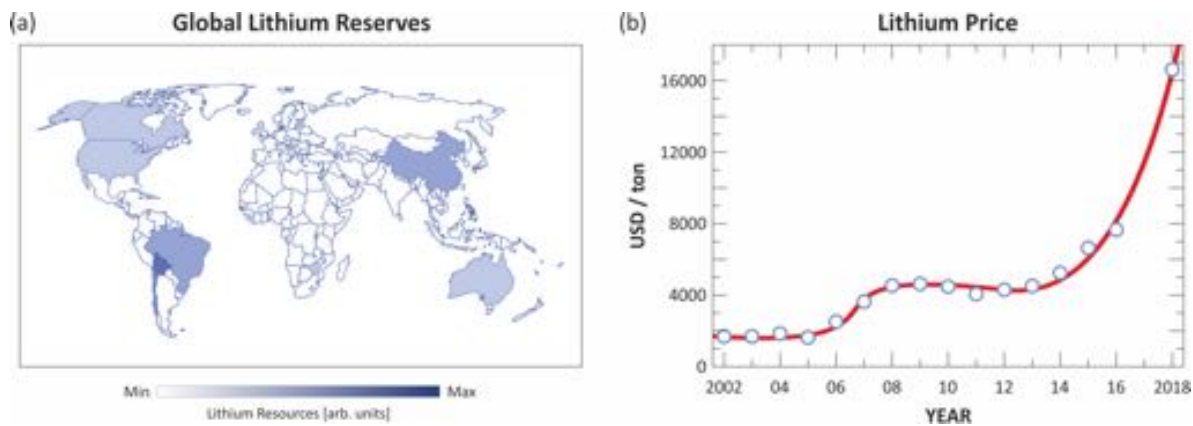

**Figure 5.** (a) Global lithium reserves showing the strongly unequal distribution with a pronounced absence of rich deposits in Europe. The map is adapted from data presented in Ref. [4]. (b) The drastic price development (increase) of lithium raw material over the last few years. The data is extracted from Ref. [5].

### 3.2. Sodium-Ion Batteries (Na)

With the aim of avoiding a monopolistic Li-based society, with the associated problems outlined above, mirroring the vulnerability to oil mentioned by John Goodenough, it would be ideal to find a viable complement, a parallel alternative (not a replacement!), to LIBs. One of the natural options would be to simply move down one step in the periodic table of elements (see Figure 6), from lithium to sodium (Na), i.e. to the realization of Na-ion batteries (NIBs). Sodium has many advantages over lithium. For one, sodium is one of the most abundant elements in the Earth's crust (with 25,000 ppm [11]) and is therefore about 1500 times more abundant than lithium. Further, it is readily accessible to every continent either from land deposits or through the salt water in our great oceans (containing about 35,000 ppm NaCl). This is clearly also reflected in the price: Sodium is a very cheap metal costing only approximately 150 USD/ton to extract and refine. In comparison, lithium is about 100 times more expensive (currently 16,500 USD/ton). Moreover, creating contacts with NIBs is more straightforward as aluminum contacts can be used, which are cheaper than the Copper ones required for LIBs (because lithium alloys with aluminum). All in all, this makes NIBs a much more cost-efficient alternative to LIBs. Finally, the higher natural abundance, easier extraction, lower health hazards and easier recycling gives NIBs a much more favorable Environmental Impact Assessment (EIA).

Obviously lithium batteries are always most likely to provide a higher energy density because lithium is smaller than sodium. Furthermore, at the current stage of development, LIB technology has evolved much more and performs considerably better than the corresponding NIBs. Consequently, for present (and potentially also future) mobile applications, LIBs are clearly the first choice. That said, there is still considerable scope for NIB applications for stationary energy storage. This market is currently expanding owing to the changeover to more sustainable energy conversion technologies, such as solar panels and wind-power. Such energy "production" is strongly decentralized and fluctuates, thereby creating a need for cheap, large-scale and temporary energy storage either directly at people's homes (Figure 6) or within the construction of so-called smart grids [16]. For such applications, the larger volume/weight of NIBs is not important, however the lower price and favorable EIA are undoubtedly extremely advantageous.

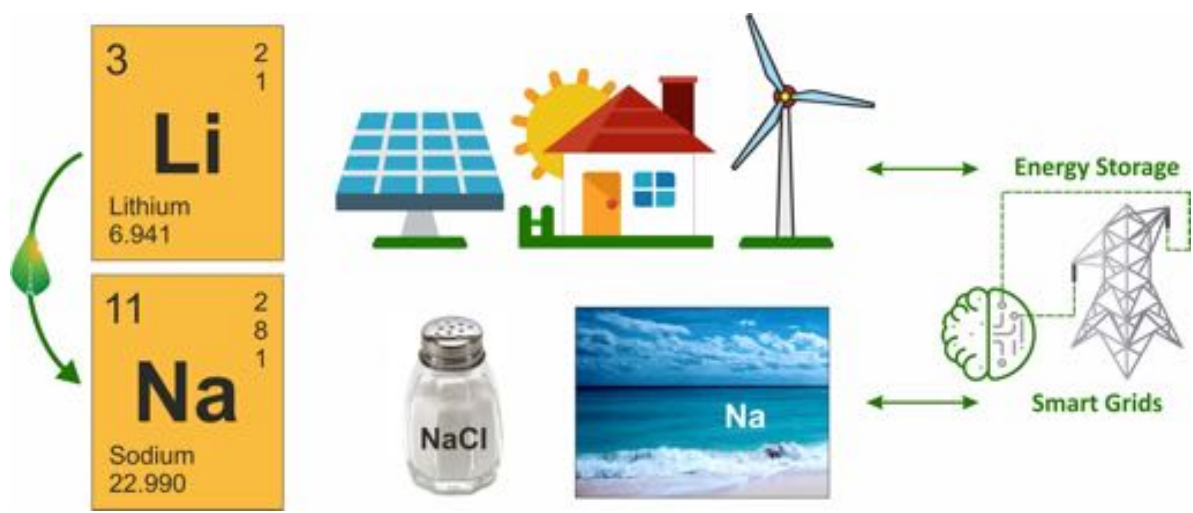

**Figure 6.** Schematic view of how moving from Li to Na based energy storage allow us to realize both local and decentralsed energy "production" (from e.g. solar panels and wind power) as well as augmenting smart grids. It should be noted that such applications are possible even with today's existing NIB technologies. Credit Martin Månsson,

*3.3. Advanced Materials Characterization: State-of-the-Art Large-Scale Facilities*

For a paradigm shift to be achieved in the field of energy storage, and, for NIBs, especially where mobile batteries are concerned, a new generation of energy materials is needed. To attain this, an understanding of the underlying microscopic mechanisms for energy conversion and energy storage is crucial. No one can ignore the tremendous evolution of LIBs during the last couple of decades. However, such development was achieved using mainly electrochemical methods, e.g. electrochemical impedance spectroscopy (EIS). Such methods do not per se yield intrinsic material properties, but rather, are a measure of device performance on a macroscopic level. To take the next step in this field and acquire true intrinsic materials properties, more advanced experimental techniques will have to be utilized: To access materials' and/or devices' structural as well as dynamic properties (Figure 7) down to an atomic scale, state-of-the-art large-scale experimental techniques, e.g. synchrotron, neutron and muon sources, are the ultimate tools. Here, Sweden is currently making unprecedented investments in large-scale research infrastructures with the recently inaugurated MAX IV [17] synchrotron facility as well as the current development and construction of a world-leading neutron facility with the European Spallation Source (ESS) [18], [19]. This will give Swedish researchers ideal opportunities to conduct leading-edge research on, e.g., sustainable energy materials within the near future.

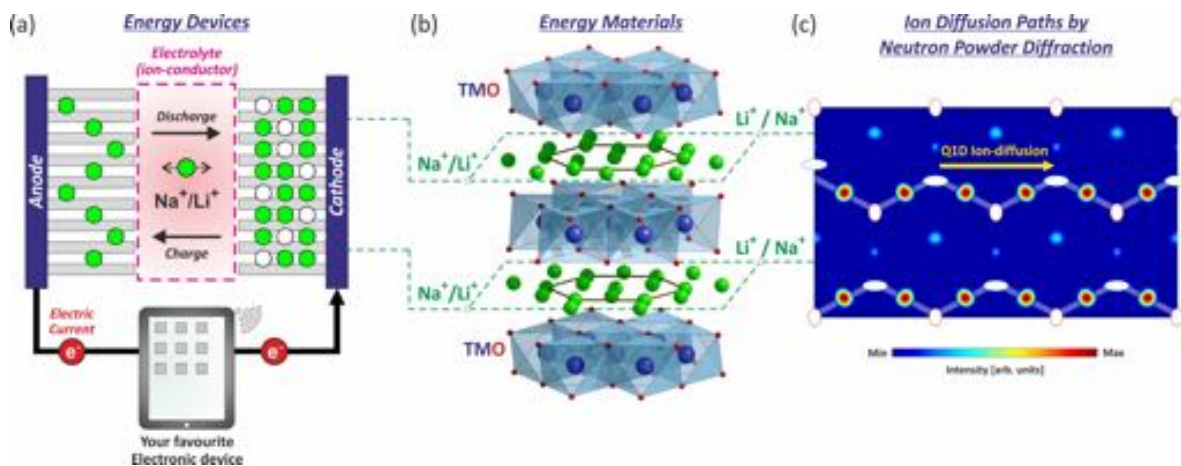

**Figure 7.** (a) Schematic description of energy storage in rechargeable batteries (b) Archetypical layered battery cathode material showing the metal-ion planes (e.g. $Na^+$ or $Li^+$) sandwiched between transition-metal-oxide (TMO) planes. (c) Quasi-one-dimensional (Q1D) ion diffusion paths in a Na-based battery material revealed by advanced Fourier analysis of neutron powder diffraction data [20]. Credit: Martin Månsson.

The ideal starting point for the development of NIBs is to use the recent advances in LIBs as a platform and conduct comparative studies on lithium-based materials and their sodium analogues. Here, our ongoing research project studied the sodium analogue of the archetypical battery cathode material $Li_xCoO_2$, i.e. $Na_xCoO_2$. By using high-resolution neutron powder diffraction (NPD) we revealed a two-step "melting" of the Na-ion planes (see also Figure 7(c)), involving an intriguing crossover from 1D-to-2D Na-diffusion [20]. Further, it is evident that the onset and evolution of ion-diffusion is intrinsically linked to a series of subtle structural transitions which unlock the diffusion pathways. Finally, by using quasi-elastic neutron scattering (QENS) to study the Na-ion dynamics [21] we confirm that the structural changes are directly linked to ion-diffusion. Neutron scattering [22] is one of the most versatile experimental methods with which to study atomic (and magnetic) structures, as well as dynamics. QENS (quasi-elastic neutron scattering) is a specialized application within inelastic neutron scattering (INS) that focuses on low-energy excitations (less than about 2 meV) close to the elastic line (Figure 8 see (a)). By studying the momentum (Q) dependence of the quasi-elastic broadening ($\Gamma$), it is possible to reveal intrinsic details of the ion-diffusion process (Figure 8 (b)), making QENS a very powerful method. However, since QENS requires very high energy resolution, it is, unfortunately, very slow and requires large sample volumes.

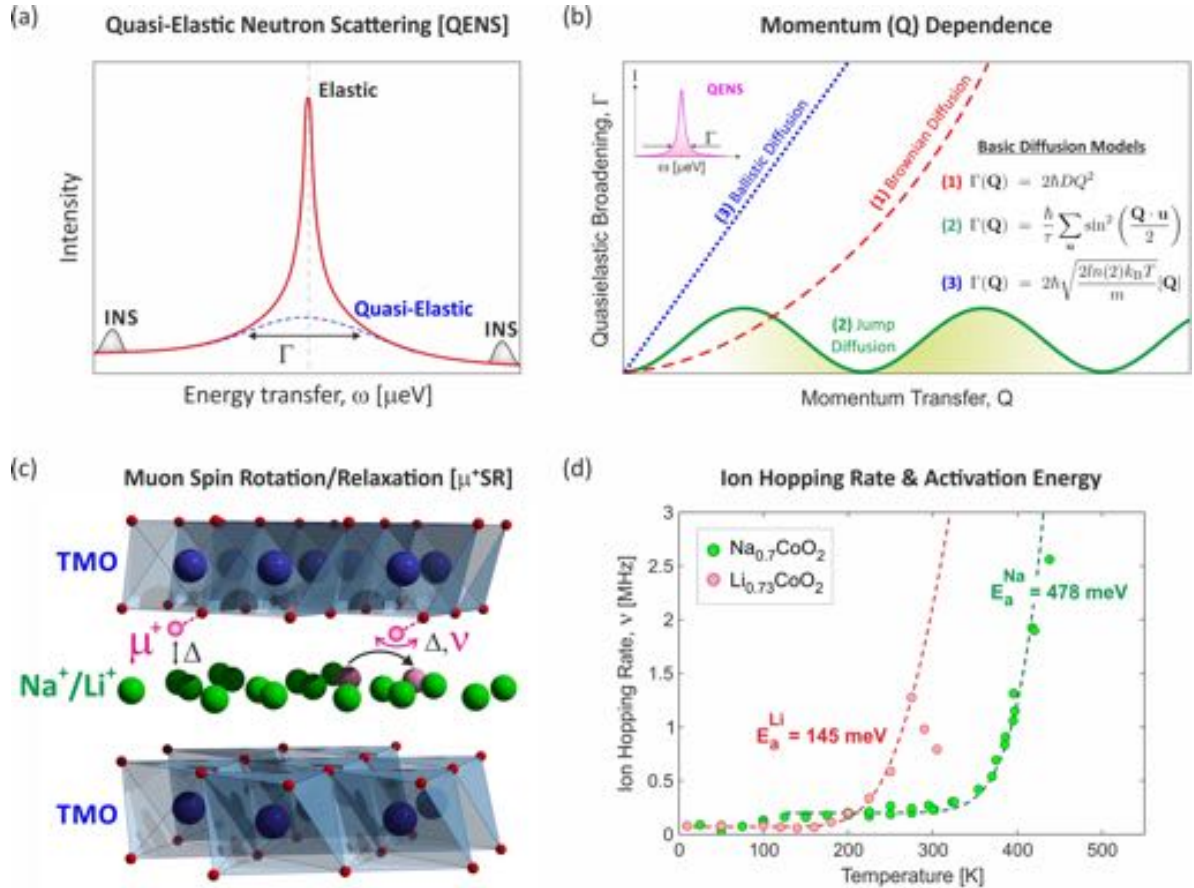

**Figure 8.** Ion-diffusion studied by (a-b) neutrons and (c-d) muons. (a) Quasi-elastic neutron scattering (QENS) concerns low-energy excitations where the temperature dependent line-broadening ($\Gamma$) shows the onset of ion motion and/or diffusion. (b) By studying the momentum dependent $\Gamma(Q)$ it is possible to discern the different types of ion diffusion process. (c) Schematic view of how the muon will feel the static or dynamic nuclear field from, e.g., Li or Na inside a layered battery material. (d) The temperature dependence of the ion hopping rate $\nu(T)$ reveals the diffusion constant $D_{ion}(T)$ and, thereby, the activation energy ($E_a$) of the diffusion process. Credit Martin Månsson.

With the intention of filling the experimental void between EIS methods and QENS for ion dynamics studies, we have developed a novel method that utilizes the muon spin relaxation/rotation ($\mu^+$SR) technique∗ [23] to probe the microscopic self-diffusion constant in a straightforward manner [24], [25]. Muons are spin-polarized (S = $\frac{1}{2}$ ) particles with a very large gyromagnetic ratio $\gamma/2\pi = 135.5$ MHz/T. As a result, when implanted into a material under zero external field (an advantage over other techniques, e.g. NMR), muons act as an extremely sensitive (fraction of a Gauss), and local, probe of static as well as dynamic magnetic/dipole fields. For a battery material in a paramagnetic (PM) state, implanted muons will mainly feel the nuclear magnetic dipole moment. By performing zero- and weak longitudinal-field (ZF; LF) measurements, it is possible to decouple the magnetic and nuclear dipole interactions. If the metal ions are not diffusing, the nuclear field is static and the $\mu^+$SR signal is dominated by the field-distribution width ($\Delta$). However, if the surrounding ions starts to diffuse, the muons will additionally detect a dynamic contribution i.e. the field-fluctuation rate ($\nu$) (see also Figure 8 (c)). In energy materials, the field-fluctuation rate can in many cases be directly translated into an ion hopping rate, providing access to the important temperature dependence of the ion diffusion constant, $D_{ion}$(T) and thereby also the activation energy ($E_a$) of the diffusion process [Fig. 8(d)]. After presenting $\mu^+$SR as a novel and optimal probe of ion dynamics [24], [25], our method has been applied to a wide range of both LIB and NIB materials over the past 10 years [26], [27], [28], [29], [30], [31], [32], [33], [34]. In addition, the method has proved to be extremely valuable for studying other dynamic processes in, e.g., hydrogen storage materials [35], ion conductors [36] and confinement materials for long-term storage within the field of nuclear waste management [37].

An interesting, versatile and powerful application of our $\mu^+$SR method is to apply it in combination with the currently available low-energy $\mu^+$SR (LEM) [38] or future ultra-slow muon microscope (US$\mu$M) [39] techniques. With this method it is possible to tune the kinetic energy of the muons, i.e. their implantation depth into the sample within the range $d_{impl}$ 5 – 300 nm. As a result, it is possible to perform depth resolved studies of ion diffusion in each of the individual components and their interfaces to, e.g., a thin film solid-state battery. The $\mu^+$SR and the related $\beta$NMR [40] technique are unique in being able to probe ion-diffusion across the interface. During the last couple of years our team has conducted the very first investigations of thin film battery materials using these techniques [41], [42]. The results have provided a novel and detailed insight into the ion-diffusion mechanisms in these compounds using neutron scattering as well as $\mu$+SR. Such knowledge and understanding now allows us to actively consider tuning energy related materials on the atomic level. This can clearly improve the materials' functional properties and, by extension, enhance device performance.

### 3.4. Concluding remarks

It should be clearly stated that this transcript should not be interpreted as a statement against LIB technology; LIBs are indeed a very efficient and useful technology that will most likely always exhibit a higher energy density than NIBs. Rather, the aim is simply to emphasize that building a modern energy society based solely on one (and only one) technology might cause an unwanted monopolistic situation such as that already experienced with oil, alluded to previously. Hence, where it is technically possible, it would be highly favorable to find an alternative to LIBs as a complement, rather than replacement form of energy storage. NIBs could be a viable alternative (for stationary storage), which would be politically, economically and environmentally advantageous owing to the ready availability of sodium worldwide. Key concepts for resolving our current energy problem are diversity and dissemination, i.e. "one technology will not save us, but many working together could". Further, to take the next necessary steps for a new generation of energy devices, a novel range of energy materials needs to be developed, and here state-of-the-art large-scale experimental facilities will be essential. With that said, further options in parallel with LIBs and NIBs are needed, where e.g. hydrolysis, hydrogen storage and fuel-cells are important technologies to consider. But, that is beyond the scope of this short interjection.

## 4. Will 3D printing be used to produce life? by Anthony Atala

It is often said that today's science fiction becomes tomorrow's science. With the advent of bioprinters, scientists and the public alike are pondering just how far the technology can advance. In the 2015 film Avengers: Age of Ultron, a "regeneration cradle" is used to create a new and more powerful body for a super-villain. Is this where today's research into bioprinting could lead? Is it possible that sophisticated 3D printers will one day be used to produce life? Currently, 3D bioprinting is being explored as a way to meet the demand for engineered tissues that has risen rapidly due to the limited availability of donor tissues and organs for transplantation. Three-dimensional (3D) printing technology shows promise for creating complex composite tissue constructs through precise placement of cell-laden hydrogels in a layer-by-layer fashion. We have worked for more than a decade to develop a system that deposits cell-laden hydrogels together with synthetic biodegradable polymers that impart mechanical strength, thereby overcoming previous limitations on the size, shape, structural integrity and vascularization of bioprinted tissue constructs [43]. The printer was demonstrated by fabricating human-scale mandible bone, ear-shaped cartilage and organized skeletal muscle, see Figure 9.

Of course, structural integrity is only one of the challenges in engineering tissues for the human body. Another is how to supply the structures with oxygen until they develop a system of blood vessels after implantation. It has been well established that the maximum nutrient diffusion distance for cells to survive without vascularity is ~100–200 $\mu$m. The creation of cell constructs larger than this scale requires vascularity. Several approaches have been used to promote mass transfer of nutrients and oxygen in engineered tissues, including growth factors that stimulate angiogenesis. The printer we developed allows for the use of microchannels with a porous lattice design that facilitate nutrient and oxygen diffusion, The bioprinted bone, ear and muscle constructs implanted in vivo showed evidence of vascularization without necrosis and the muscle constructs showed the presence of neuromuscular junctions. Evaluation of the characteristics and function of these tissues pre-clinically in vitro and in vivo showed tissue maturation and organization that may be sufficient for translation to patients.

With further development, this technology may produce clinically useful tissues and organs that incorporate multiple cell types at precise locations to recapitulate native structure and function. Future development is being directed to the production of tissues for human applications, and to the building of more complex tissues and solid organs. The science shows that 3D bioprinters can clearly produce living tissues, but can they produce life itself? In the near-term, the likely answer is that reproductive organs produced by bioprinting will allow some patients currently unable to conceive or carry a fetus to do so. We are currently working to fabricate a variety of reproductive organs using the 3D printer, including testicles, ovaries, vagina and uterus.

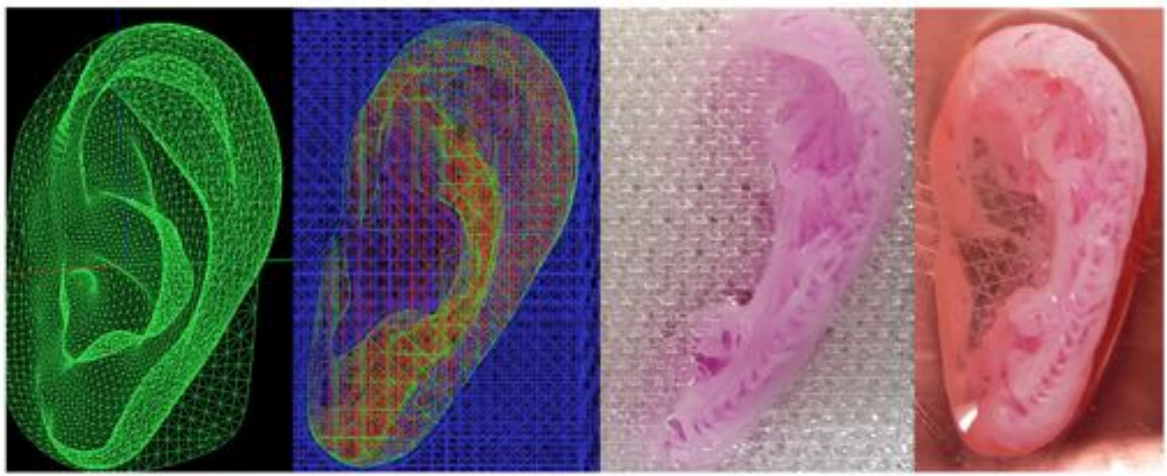

**Figure 9.** Using data from CT and MRI scans, a 3D bioprinter has the potential to "tailor-make" tissue for patients. For a patient missing an ear, for example, the system could print a matching structure. Credit: Wake Forest Institute for Regenerative Medicine.

Where do these reproductive technologies stand today? A hand-fashioned version of the vagina has already been successfully implanted in patients. The structures were implanted in a small group of women born with the Mayer-Rokitansky-Küster-Hauser (MRKH) syndrome, a rare genetic condition in which the vagina and uterus are underdeveloped or absent [44]. The organs showed normal structural and functional variables with a follow-up of up to 8 years. Work on other reproductive organs, such as the uterus, testicles and ovaries, is currently at a pre-clinical level. To engineer testicle organoids (Figure 10), we use spermatogonial stem cells. One clinical application is to re-implant the cells into young men who were made sterile from childhood cancer treatments. Another is to engineer testicular organs for men who've lost their testicles due to cancer or injuries. In vitro, the engineered organoids can secrete male hormones and have the potential to produce sperm, providing function similar to a normal organ. Our ovary research focuses on producing bioartificial organs for hormone replacement therapy [45] and with the potential to restore fertility.

Significant advances need to occur before 3D printers can be used to implant engineered complex organs, such as the kidney and liver in patients. A major challenge is the high oxygen requirement of these organs. It is likely that current-generation scientists will spend the remainder of their careers grappling with these challenges. And where does that leave the question of whether a 3D printer can create life? Perhaps one day, generations from now, scientists will have a regeneration cradle like in the Avengers film and will extend the technology in ways that only science fiction can envision today.

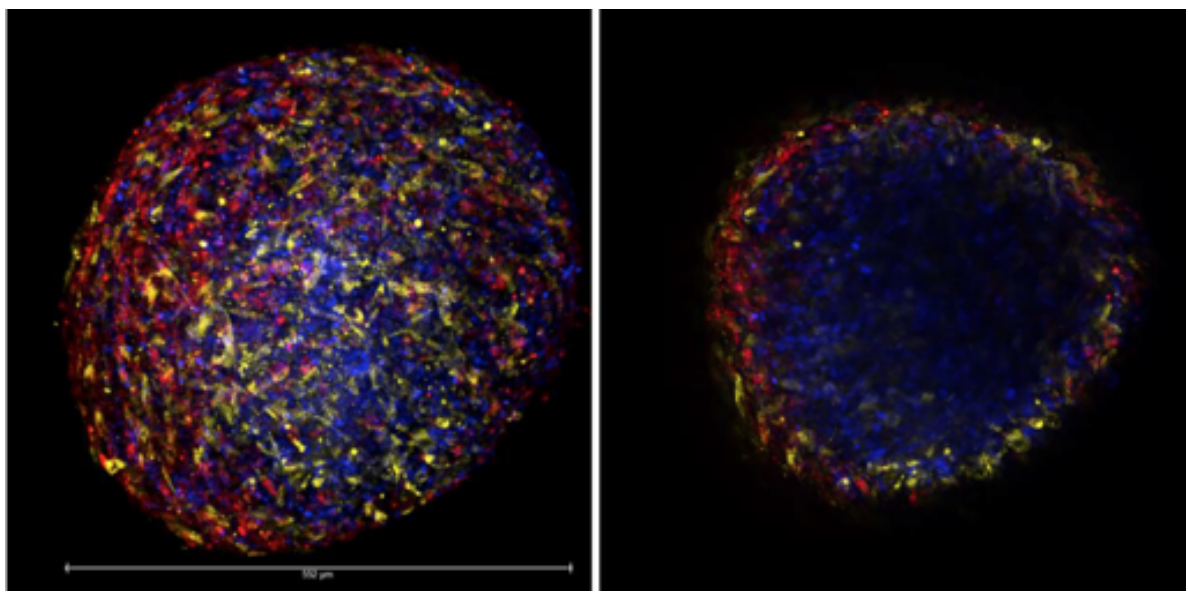

**Figure 10.** Lab-engineered testicular organoids have the potential to replace the need for hormone replacement therapy and to potentially produce sperm. Credit: Wake Forest Institute for Regenerative Medicine.

## 5. If a human were to be created atom by atom, molecule by molecule, would it behave like you and me? by Gerianne Alexander

If all the necessary biological units could be assembled to create an adult human, would this individual behave like us? Welcomed into a social group, could this new group member communicate ideas and emotions, form relationships with others, regulate internal states, and satisfy needs and wants in a socially-defined appropriate manner? Socially competent adults can do so, typically without conscious effort. However, unlike the constructed adult, we are not born fully grown. The decades from birth through reproductive maturity to emerging adulthood represent a long period of brain development within a particular social environment [46], an enriched maturational process that is argued to be necessary to support the cognitive and social competencies that allow the adult to function well in a society or culture [47], [48]. Our constructed adult would provide a powerful test of that dominant hypothesis.

### 5.1. Talking without speaking and hearing without listening

Infants are able to detect language specific sound patterns and use probability statistics to link phonemes to form the building blocks of words [49]. Early exposure to the native language also influences the perception of speech – by 12 months of age, infants lose their earlier ability to discriminate the distinct units of sound that make up all languages [50]. Infants next progress from one-word, two-word, to three-word sentences, eventually acquiring the vocabulary and syntactic structure of the native language [51]. In contrast, nonhuman primates with extensive language training show significant deficits in vocabulary and language syntax, suggesting that the acquisition of these essential aspects of language is a unique capacity of the human brain [52]. However, the similar language training outcomes of children deprived of exposure to language during early life [53], indicate that this unique capacity for language is greatly reduced in later development [52].

That being the case, our constructed adult would have a limited capacity for speech. If, however, early experience and not time defines the critical period for language learning [49], then for this adult brain, devoid of earlier language processing, the infant's predictable path to language might still be available for travel.

## 5.2. Take my arms that I might reach you!

Beyond language, effective social communication requires an ability to convey and recognize emotional states. Facial expressions associated with basic emotions (e.g., anger, fear, sadness) are displayed across development, across cultures, and even in individuals blind from birth [54], [55]. Preferences for body characteristics that signal reproductive maturity and fertility, such as broad shoulders in men and small waist to hip ratios in women, are also found across cultures [56], [57]. Congenitally blind men, feeling the hips and waist of female mannequins, also prefer the female with a smaller waist [58]. Together, these findings suggest that our constructed adult will be biologically prepared to recognize basic emotions and value physical traits necessary for survival and reproductive success. However, other indicators of emotional states such as body posture or hand signals are culturally specific. A thumb and finger joined might indicate "all is OK" in one culture, but be viewed as an obscene gesture in another. Healthy adult friendships and intimate partnerships are thought possible only because we hold mental representations of our self in relation to others, an "internal working model" acquired through our early attachments to caregivers [59]. Similarly, the often gendered peer relations in childhood and adolescence [60] result in the internalization of social-sexual scripts that dictate the rules of courtship and bonding [61]. Our constructed adult may seek to satisfy needs for friendship and sexual intimacy. However, being effectively blind to the social signals from others inviting approach or avoidance and lacking social scripts for engagement, the necessary guides to appropriate social interactions that permit realization of these goals would be unavailable to our constructed adult.

## 5.3. In restless dreams I walk alone

Perhaps central to all, socially competent adults have a strong sense of self – a knowledge of the characteristics of things and people that they like, what they want to achieve, and an understanding (not necessarily accurate) of their strengths and our weaknesses. Much of this self-knowledge that directs our behavior comes from an identification and comparison with other members of our social group [63]. The cliques associated with adolescence and emerging adulthood, as an example, are made up of individuals who share common interests, modes of dress, and ways of acting. Not being associated with a clique in adolescence is associated with maladjustment [64], suggesting the importance of group affiliation for the newly formed adult.

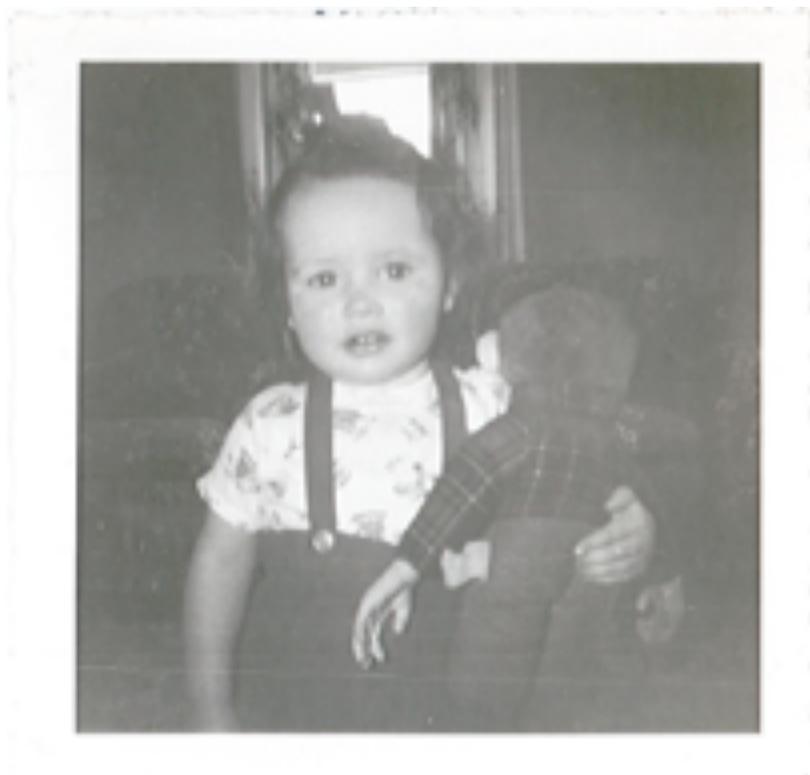

**Figure 11.** Shown here at a young age, Gerianne Alexander's primate research has revealed a sexually differentiated choice of toys in non-human primates suggesting that such preferences observed in young children might have arisen prior to the emergence of a distinct hominid lineage [[62] ]. Credit: Gerianne Alexander.

Gender is one of the most salient social group categories. Most adults have a gender identity, a sense of being male or female that begins with the self-labeling of gender around the age of three years of age and the construct is subsequently elaborated by identification with group members sharing the same gender label [65]. Gender identity, thus, supports a network of learned associations between gender label and gender-typical behaviors called gender schemas that guide our behavior so be gender-consistent. Children's strong gender-linked toy preferences (e.g., dolls, trucks, tea sets, tools), for example, provide experience with miniature replicas of objects associated with socially prescribed roles for women and men [66]. However, monkeys and young infants presumably without a self-awareness of gender also show "gender-typical" preferences for trucks and dolls [62], [67], suggesting that even in the absence of gender socialization, biological influences (perhaps on temperament or visual preferences) [68] might nudge our adult to affiliate with one gender group over another.

### 5.4. Planting in my brain

How quickly could our constructed adult adapt? There is clearly plasticity in the adult brain [69] London cabdrivers, able to navigate through the complex roads of the city, have larger brain areas specialized for spatial navigation [70]. So, like us, the behavior of the fully formed adult will be shaped by experience - and the positive or negative consequences of behavior will move the constructed adult towards a greater approximation of the socially competent individual. Consider, however, the less than optimal success of interventions aimed at overcoming the deficits in social communication associated with autism spectrum disorder [71] or those aimed at increasing a limited capacity to form relationships in adults because of impairment in emotion regulation and emotion recognition [72]. If our constructed adult without a past could easily become a socially competent adult, then that outcome would suggest that the contribution of biological factors to social competency is far greater than we believe. Yet, lacking a long past of personal experiences, the constructed adult will still differ from us in a fundamental way. And without the possibility of mental time travel to the self in the past, what would inform the mental travel to the possible self of the future [73]?

## 6. What is consciousness, and do we have free will? by Roland E. Allen and Suzy Lidström

Much of what has been written or spoken about consciousness is of dubious value, and even the best contributions have sometimes been received with confusion. Here we wish to establish several clarifying points and then briefly describe a simple formal model in the spirit of one modern interpretation [74], [75].

At the end, however, two questions will still remain: What are the detailed physical processes that correlate with inner consciousness, and is there really a "hard problem" [76] that lies beyond normal scientific explanation? The second question can be rephrased as follows: Is it at all possible, even in principle, that we might someday be able to explain why our inner experiences are what they are, as we directly feel or sense them – red as red, cold as cold, pain as pain, a pleasant memory as pleasant, and a concept as an abstract generalization of the raw input from our senses?

In this short contribution it is impossible to cite the vast number of ideas, papers, and books on consciousness (and free will, a related but separate subject), but some of the most relevant and prominent scientific aspects will be referenced in a forthcoming paper [77].

Our first point is this: In the enormous literature on this subject, confusion often arises from misuse or misunderstanding of language. The need for a careful analysis of language has long been emphasized, for example by the Cambridge and Austrian philosopher Ludwig Wittgenstein and his followers, and it is particularly obvious in the present context.

In order to avoid confusion, we therefore adopt the convention of distinguishing different meanings of a word with subscripts. By consciousness$_O$ we mean the state available to an external observer of nature and thus to scientific description – presumably a set of neuronal processes in the brain that is accessible to scientific probes. By consciousness$_P$ we mean the internal experience of a human participant within nature which contains the "qualia" of redness, coldness, etc. Consciousness$_O$ will presumably correlate with consciousness$_\mathbf{P}$ but the two concepts are quite distinct.

Similarly, there are different forms of knowledge. When you, as a human participant, experience pain, you know what pain is in one sense. But if you, as a human observer, note the way in which pain is expressed, or even can identify in great detail the neural correlates of pain in a human brain, you know what pain is in a very different sense. These are respectively the inner feeling pain$_P$ and the observed phenomenon or scientific description pain$_\mathbf{O}$. In general, there is an inner know$_\mathbf{P}$ and an outer know$_\mathbf{O}$.

In standard usage, principally in philosophy, "quale" has only the meaning quale$_\mathbf{P}$ – "a property as it is experienced as distinct from any source it might have in a physical object" – and an attempt to use this word in the sense of a quale$_\mathbf{O}$ will tend to produce confusion. But if the scientifically observable neural correlates of qualia$_P$ are determined and called qualia$_\mathbf{O}$, the two terms should still be recognized as completely distinct.

A second source of confusion is the implicit – and incorrect – assumption that this distinction between an outer scientific description and the inner reality of nature is limited to human consciousness. In fact, this distinction applies to all of nature, from humans to bats [78], bees [79], stars, protein molecules, computer memories, and all the rest of the universe in its smallest to largest aspects. The only inner reality (reality$_\mathbf{P}$) we can directly experience is that associated with neuronal processes in our brains. For everything else in the world around us – other conscious life, life without consciousness, nonliving matter, and immaterial fields – we have only the representations provided by those processes. Their inner reality is not directly accessible to us.

The statement that all of nature has an inner reality (what Kant would call the Ding an sich) does not, of course, mean that all of nature is conscious. The model of [77], in the general framework of e.g. references [74] and [75] provides an interpretation of what consciousness is and what it requires, with the implication that only a very tiny part of the substance of the universe possesses consciousness.

The main point, again, is that we can directly know only one reality – our inner experiences. These are our representation of reality. When we see an object as red$_\mathbf{P}$ (the inner experience of red) it is because specific cone cells in the retina have been stimulated by electromagnetic waves near a specific wavelength, and have themselves stimulated the neural processes that correlate with red$_\mathbf{P}$, which might be called red$_\mathbf{O}$. (Of course, the light of this wavelength, the object emanating the light, etc. can also be called red, but this is a familiar ambiguity.)

Similarly, a bat "seeing" moths with high-frequency sound or a bee with compound eyes experiences a reality that might be called bat-perception$_P$ or bee-perception$_\mathbf{P}$, which we can never know directly, but we can in principle describe scientifically as bat-perception$_\mathbf{O}$ or bee-perception$_\mathbf{O}$. In every case, $X_P$ is the inner reality of nature whereas $X_O$ is the scientific representation of that reality.

Every description in science is somewhat analogous to a black and white map of the coloured true terrain of nature. The maps can become more and more refined as science progresses, but even an ultimate and perfectly faithful description would not be the same as the inner reality itself.

One source of confusion is that a human being can assume the role of either observer or participant. But the roles are clearly distinct. Even if one were to observe images of the correlated neural activity while experiencing the sensation of redness, the red$_\mathbf{O}$ in the brain images on the screen and the red$_\mathbf{P}$ of inner experience are quite distinct.

In this vein of clear thinking – or philosophy as a prelude to science – a question arises: Does David Chalmers' "hard problem" of consciousness actually exist?

We will first present a relatively narrow argument that it does not exist, for the same reason that "the sound of one hand clapping" does not exist. I.e., if "problem" and "knowledge" have their usual meanings, there is in fact no "hard problem" left if a scientific description of consciousness can be achieved. Within this frame of discussion, "the hard problem" (as it is defined) is a contradiction in terms or logical impossibility, like "square circle". It would then follow that phrases like "the hard problem" or 'the sound of one hand clapping" can inspire unconventional thinking (as well as artistic endeavors like the recent Tom Stoppard play), but are fundamentally nonsensical.

After this narrow argument, however, we will move to a broader perspective and consider the possibility that "problem" and "knowledge" may have more general meanings, which cannot yet be clearly stated, but may result from a deepened future understanding of nature – which presumably means deeper physics.

In the narrow frame of discussion, the "hard problem" is an example of the confusion that results from improper use of language, in this case the word "problem". There are highly nontrivial and potentially deep problems involved in the pursuit of scientific understanding of consciousness – i.e., consciousness$_\mathbf{O}$. But if such understanding is ever achieved, there will be no extra problem left over. I.e., the word "problem" has only the meaning problem$_\mathbf{O}$ and there is no problem$_\mathbf{P}$ for the following reason:

Suppose that we do attain a complete scientific understanding of mental processes. Then we will know$_\mathbf{O}$ what the sensations of red, cold, etc. are, what thoughts and memories are, what emotions like fear and happiness are, and what consciousness is. But we will also continue to know$_P$ what these things are, through direct experience. And the only possible link between the two kinds of knowledge is the one provided by the scientific correspondence between them. There is then no further knowledge to be had, and no "hard problem" left over.

Furthermore (still within this narrow frame of discussion), if we extend our attention to phenomena outside human consciousness (paralleling the progress of science in removing humans from their central position in the universe), we can only know$_O$ these phenomena. But it is logically impossible for us to know$_P$ anything outside our own neural processes. So again there is no further knowledge to be had, and no "hard problem" left.

Let us now, however, move to a broader frame of discussion, with potentially broader meanings of "know", "understand", "problem", and even science itself. We want to avoid the trap into which linguistic philosophy has sometimes fallen, of trying to abolish true philosophical problems by overly restrictive constraints on language. Precision of language is important, but well-considered extensions of language are permitted.

Any object in nature can be correctly described at many different levels – e.g., a human being can be described as a person, a collection of organs, a collection of cells, a collection of molecules, a collection of fundamental particles, or a collective excitation of quantum fields. In present-day physics, this last description is the most fundamental available, and even speculative theories such as string theory and loop quantum gravity are qualitatively no different from quantum field theory. We do not know, however, where the most fundamental physics will take us in understanding during the coming centuries, or what form it will take.

It follows that future physics may in fact be able to address the "hard problem", and may be able to explain why redness, pain, and all the other aspects of inner consciousness take the forms that we directly feel and experience.

In this sense, the "hard problem" has the same status as the problem of "why is there something rather than nothing?" [80] We do not know whether these problems can be answered by – or even have meaning for – creatures like ourselves who are embedded within nature.

But even if this hard$_P$ question cannot possibly be addressed within present-day physics, the scientific hard$_O$ question is enormously interesting: What are the detailed physical processes that correlate with inner consciousness?

This question will be addressed in a paper where a simple formal model will be proposed [77], which is meant to be a kind of template for organizing the almost overwhelming complexity of neuronal processes. Here we just say that the model appears to be novel, in the sense that consciousness is addressed at the levels of both biology and physics, and that it is interpreted as a collective excitation of neurones or quantum fields which spans all the relevant parts of the brain.

The model is thus quite different from those based on only a localized structure in the brain, information per se, quantum coherence, etc. In the model of Ref. [77], consciousness is roughly analogous to global vibration of a molecule which consists of weakly bonded molecular fragments, with sensations, memories, motor control, etc. roughly analogous to vibrations of these individual fragments. Consciousness and each of its components are thus modeled as collective neuronal modes (or collective modes of quantum fields from a physics perspective), in an extraordinarily complex interacting hierarchy.

The change of emphasis from local structures to a superposition of collective modes is analogous to the change from atomic coordinates to normal mode coordinates in a molecule or material. Delocalized excitations can be just as real as localized excitations, as demonstrated by phonons or plasmons in a solid.

We end by considering the separate but related topic of free will. It should first be realized that quantum mechanics per se is just as deterministic as classical mechanics, and that human beings – as physical systems within nature – are described by the deterministic evolution of a state $\psi$ in time $t$ according to the Schrödinger equation

$$i\hbar\frac{\mathrm{d}\psi}{\mathrm{d}t} = H\psi$$

where $H$ is the Hamiltonian operator. (Invoking the indeterminism of quantum measurements in this context – the issue of free will – represents an extreme limit of confusion.) So human beings would not have free will from the perspective of a God-like observer who is able to make predictions with a precision that will be forever impossible for any real observer, even with the most advanced technology imaginable.

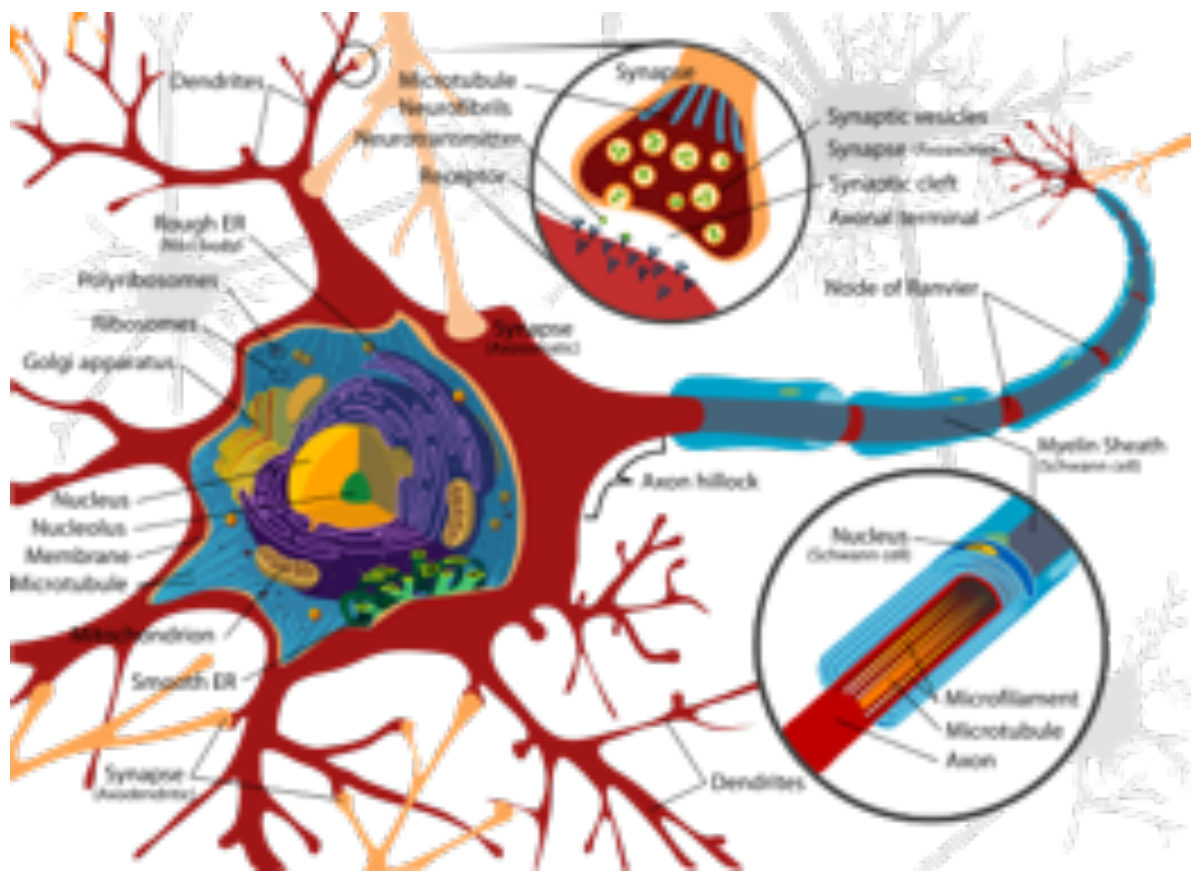

**Figure 12.** The complexity of even a single neurone. Credit: Mariana Ruiz Villarreal, Wikipedia Commons

However, one should again make the distinction between free-will$_P$ and free-will$_O$, and there is a simple proof by contradiction that a participant within nature can have

free-will$_P$, which is related to the self-reference paradox discussed by Bertrand Russell and others (extending at least back to Epimenides, and forward beyond Gödel): Within set theory (for example), one is not allowed to define a set which implicitly refers to itself. For the same reason, you cannot consistently predict your own behaviour or receive a valid external prediction of your behaviour. If a God-like being were to tell you how you will behave one minute in the future, you can (and probably will) perversely choose a completely different course of action.

You therefore have free-will$_P$ (within the constraints imposed by outside circumstances) – and this is precisely what is meant by the normal usage of the phrases "free will" and "free to choose". (There is, of course, the experimentally accessible question of what unconscious neural processes precede your conscious awareness of having made a decision, but this is a separate issue.)

But again, in summary of the earlier discussion above, we are still confronted by two profound questions: What is the scientific explanation of consciousness, and is there really a "hard problem" that lies beyond normal scientific explanation? The first question will require further advances in experimental techniques for determining how the tens of billions of neurones, like that in Figure 12, interact through their intricate connections and networks. The second may require new physics.

## 7. Can quantum techniques tell us about the dynamics of single molecules in their native state? by Warwick Bowen, N.P. Mauranyapin and L.S. Madsen

Motor molecules are the nanoscale machinery of life. They are responsible for DNA transcription, replication and recombination, transport of nutrients within and between cells, the release and storage of energy to power chemical reactions, along with many other processes (see e.g. [81]). Life as we know it simply could not exist without them. This motivates efforts to understand how these motor molecules function, move, and respond to their environment. Ideally, such investigations would be performed at the level of the dynamics of single motor molecules in their natural state and operating in their native environment. However, this is an exceptionally challenging task owing to the small size-scales involved – well beneath the diffraction limit of light (see Figure 13).

The difficulty can be seen by a simple example. Consider a protein molecule in water illuminated by light. The first-order interaction is that of elastic scattering – the optical electric field polarizes the molecule, and this rapidly oscillating polarization radiates light via dipole scattering. Let us assume that the light is focused to a tight waist of width roughly equal to its wavelength $\lambda$, and that the molecule is spherical with radius $a$. It is straightforward to show that the fraction of incident photons that scatter from the molecule is on the order of $(a/\lambda)^6$ [87]. The scaling to the power of six means that only one in $\sim 10^{15}$ photons are scattered for near-infrared illumination of a typical protein of 3 nm radius. Leaving aside the challenging task of discriminating this very low level of scattering from background scatter, one would naively think that to detect the molecule would require at least one photon to be scattered within the detection time $t$. This introduces a quantum limit to the illumination intensity $I$ for single molecule sensing via dipole scattering:

$$I \geq \frac{\hbar\omega}{\lambda^2\tau} \left(\frac{\lambda}{a}\right)^6 \tag{3}$$

where $\omega$ is the optical frequency. If we wished to track the dynamics of our 3 nm molecule over millisecond timescales, we find that the minimum optical intensity required in an ideal scenario is around $10^{11}$ W/m². But this is several orders of magnitude above known intensities at which the light intrudes on typical biological specimens, damaging their structure, function, growth and/or viability (see e.g. [87], [88], [89], [90], [91], [92]).

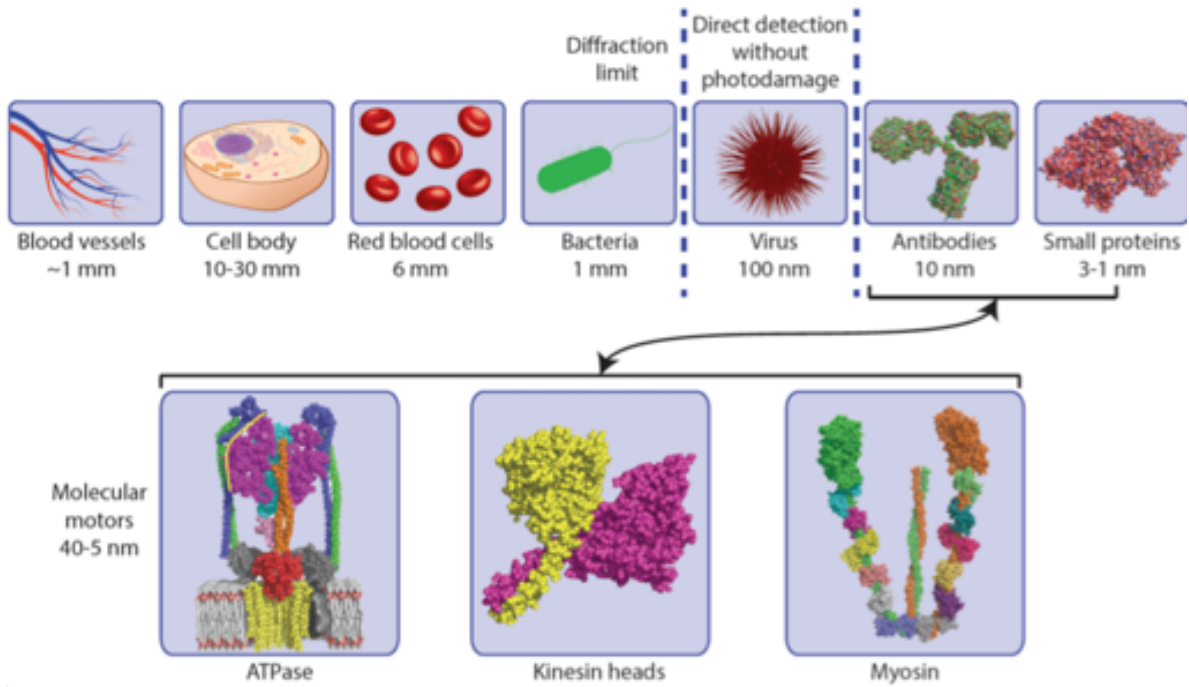

**Figure 13.** Length scales in biology: Biological entities are observable by wide field optical microscope down to the diffraction limit, which is around 200 nm, the size of organelles. To observe biological label-free specimens below this limit, other techniques have been developed based on dipole scattering of a high intensity optical field. However, photo-damage can occur at such high intensities which makes it difficult to non-perturbedly observe particles smaller than 70 nm. Many molecular motors are of a size below this limit and remain to-date unobserved in their natural state. The molecular structures of the antibody, small molecule (BSA), kinesin heads and myosin were generated from x-ray crystallography in refs. [82], [83], [84], [85], respectively. The ATPase structure is reused from [86] with permission from Springer Nature.

So, given that just looking at small biological molecules is sufficient to perturb them, and to alter their environment, how does one study their dynamics in their native state? There is, to date, no fully satisfactory answer to this question. Labels, such as microparticles or fluorescent markers are commonly used to increase the scattering cross-section [93], but these can alter the chemical environment of the molecule, and in order to enhance scattering must generally be much larger than the molecule itself, which significantly changes how it moves. Alternatively, shorter wavelength illumination – such as x-rays can be used [94], but to date these sources have proved more damaging to biological specimens than light. Another option is to use an electron beam, as in the 2017 Nobel Prize for Chemistry [95], but this requires cryogenic conditions.

Given that it is ultimately the quantization of light that limits single molecule sensing, it may be that quantum optics holds the answer. Perhaps new nanoscale probes based on quantum defects in diamond or other materials may provide the enhanced nanoscale interaction necessary to gently probe single molecule dynamics [96]. Or perhaps quantum engineering of optical fields, which has been shown to allow the detection of nanoparticles in biological specimens at intensities below the quantum limit [97], may break the deadlock. While the answer is not yet clear, what is clear is that this is a crucially important question which, if solved, may greatly advance our understanding of the machinery of life.

## 8. How can chaos be exploited in science and technology? by Linda Reichl

One might consider this a strange question given that technology is generally focused on trying to locate pockets of *order* in an otherwise *chaotic* world. Technology must always bow to thermodynamics, which is the theory of all matter governed by short ranged interactions. Thermodynamics is based on variables that emerge from a few symmetries in a microscopic world that has a huge number of degrees of freedom continually in chaotic (ergodic) motion.

Chaos is largely a classical concept. It requires a system with a continuous spectrum. In quantum systems with a discrete spectrum, one can only look for the quantum manifestations of chaos [98]. Until recently, the focus of quantum chaologists has been to locate manifestations of chaos in the natural quantum world - in the dynamics of atoms, molecules, and nuclei. However, advances in technology have now allowed scientists to build devices that are mesoscopic and even microscopic in size, so that the quantum dynamics can be shaped and controlled. In quantum devices, often the aim is to avoid the manifestations of chaos because it is accompanied by the destruction of symmetries and the entanglement of quantum states. This, in turn, leads to the spreading of the probability throughout the available quantum state space. However, this is not always a bad result.

One of the most important examples of the technological exploitation of chaos in quantum systems is STIRAP (stimulated Raman adiabatic passage), which allows control of transitions in atomic and molecular systems using slowly varying laser pulses [99]. It was shown in [100] that the STIRAP process can be analyzed in terms of adiabatically varying Floquet states that describe the atom-laser system. As the laser pulses pass through the system, they induce chaos which allows the Floquet states to undergo a Landau-Zener type avoided crossing [101], [102]. After passage of the laser pulses, the quantum system returns to its normal configuration but is left in an altered

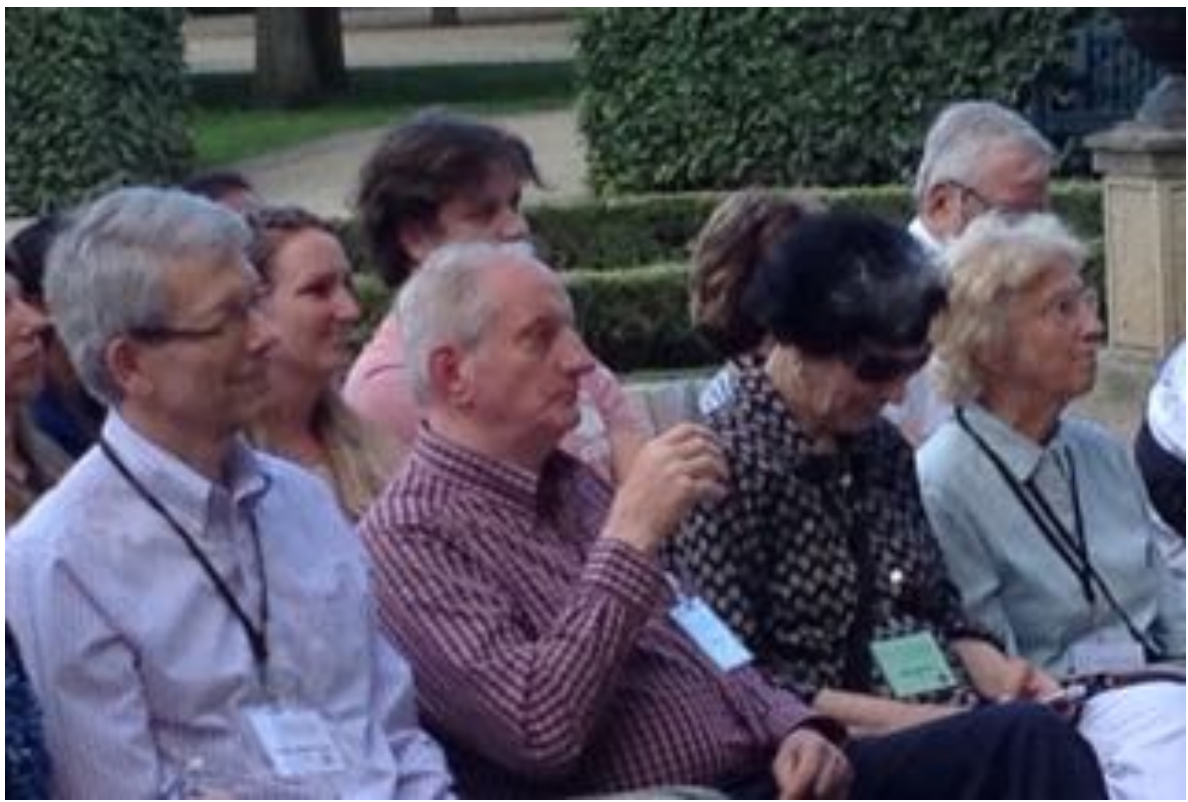

**Figure 14.** Participants enjoying the reception of the Frontiers of Quantum and Mesoscopic Thermodynamics Conference in Prague in 2017. From left to right: Steve Girvin, a leading authority on the potential experimental realisation of quantum computing, Ed Fry who did an early experiment on the validity of quantum entanglement, Debbie Fry, and Linda Reichl, Co-Director of the Complex Quantum Systems at the University of Texas at Austin. Credit: Suzy Lidström.

state - and therefore a controlled quantum transition has occurred. STIRAP is now a well defined technique with numerous applications in a variety of fields. Some of these are discussed in the recent review paper [103].

The dynamics of particle waves in small open quantum systems and of electromagnetic waves in optical microcavities have become fields of growing interest [104], because of their possible technological applications. Nöckel and Stone [105] showed that optical resonators, with internal chaotic dynamics, could be used to construct lasers with unique directional emissions. Subsequently, quantum wave dynamics in a D-shaped cavity was shown in [106] to exhibit a range of dynamic behaviors ranging from fully chaotic to integrable. Redding et al [107] then used this variable dynamics to show that an electrically pumped semiconducting chip, in the shape of a chaotic D-shaped cavity, could be used to create a multimode laser with superior full field imaging capabilities. This new application of chaotic wave dynamics in small open quantum and electromagnetic systems opens the possibilities for a range of unique quantum and optical devices.

There is another type of quantum and optical device where chaos could affect the dynamics. Control of the band structure in two-dimensional lattices is an area of growing technological importance. Two-dimensional atomic films have important electronic applications at the nanoscale and photonic crystals are being developed for light-based communication systems [108]. Thermalization of two-dimensional lattices, due to broken dynamical symmetries attributable to the onset of chaos in the unit cell of such devices, could affect the possibility of optimizing device properties by influencing and changing band structure [109].

## 9. Do Bose-Einstein Condensates of cold quantum gases have any practical applications? by Ernst Rasel

The achievement of Bose-Einstein condensation (BEC) in dilute atomic gases sounded the bell for a new area in atomic physics. Today research on quantum degenerate gases and strongly correlated systems for simulation of condensed-matter phenomena accounts for the majority of the activity in this area. The question of applications of Bose-Einstein condensates already motivated the Swedish academy when the Nobel prize was awarded to Eric Cornell, Wolfgang Ketterle, and Carl Wieman for their achievements. In the press release, the academy stated that "It is interesting to speculate on areas for the application of BEC. The new 'control' of matter which this technology involves is going to bring revolutionary applications in such fields as precision measurement and nanotechnology." [110]

More than fifteen years later, interferometry with Bose-Einstein Condensates appears to be one of the most promising pathways for future matter wave interferometry and its application in metrology, fundamental physics, and last but not least in inertial sensing and gravimetry. Matter wave interferometers with laser cooled atoms already compete with today's classical gravimeters and reach uncertainties of a few $10^{-8}$ m/s$^2$ [111], [112]. Having developed the necessary methods to achieve miniaturized sources with a high flux [113], and having tackled the detrimental influence of the mean-field energy, interferometers employing delta-kick collimated Bose-Einstein condensates [114], [115] are set to exceed the state-of-the-art in several ways. They should perform with an uncertainty reduced by at least one order of magnitude due to the better control of the atomic ensemble. The extremely low effective temperatures allow for innovative methods to coherently manipulate the atoms giving rise to new interferometers based on symmetric Bragg scattering and high-fidelity transfer of chiliad photon momenta [116] for improving state-of-the-art gyroscopes [117], [118], quantum tilt meters [118], [119], [120], 11, [121], gravimeters [115], [122] or gradiometers [123], generally, all inertial sensors. Moreover, the compactness of atom-chip based sources is preparing the way for radical miniaturization [115]. This enables space-borne sensing, where the extended free fall will improve the precision by several orders of magnitude with respect to present terrestrial interferometers. The successful creation of the first Rubidium Bose-Einstein condensate in space in January 2018 makes us curious about the future – what will be the next major step to occur in this field [124] [125]? It seems

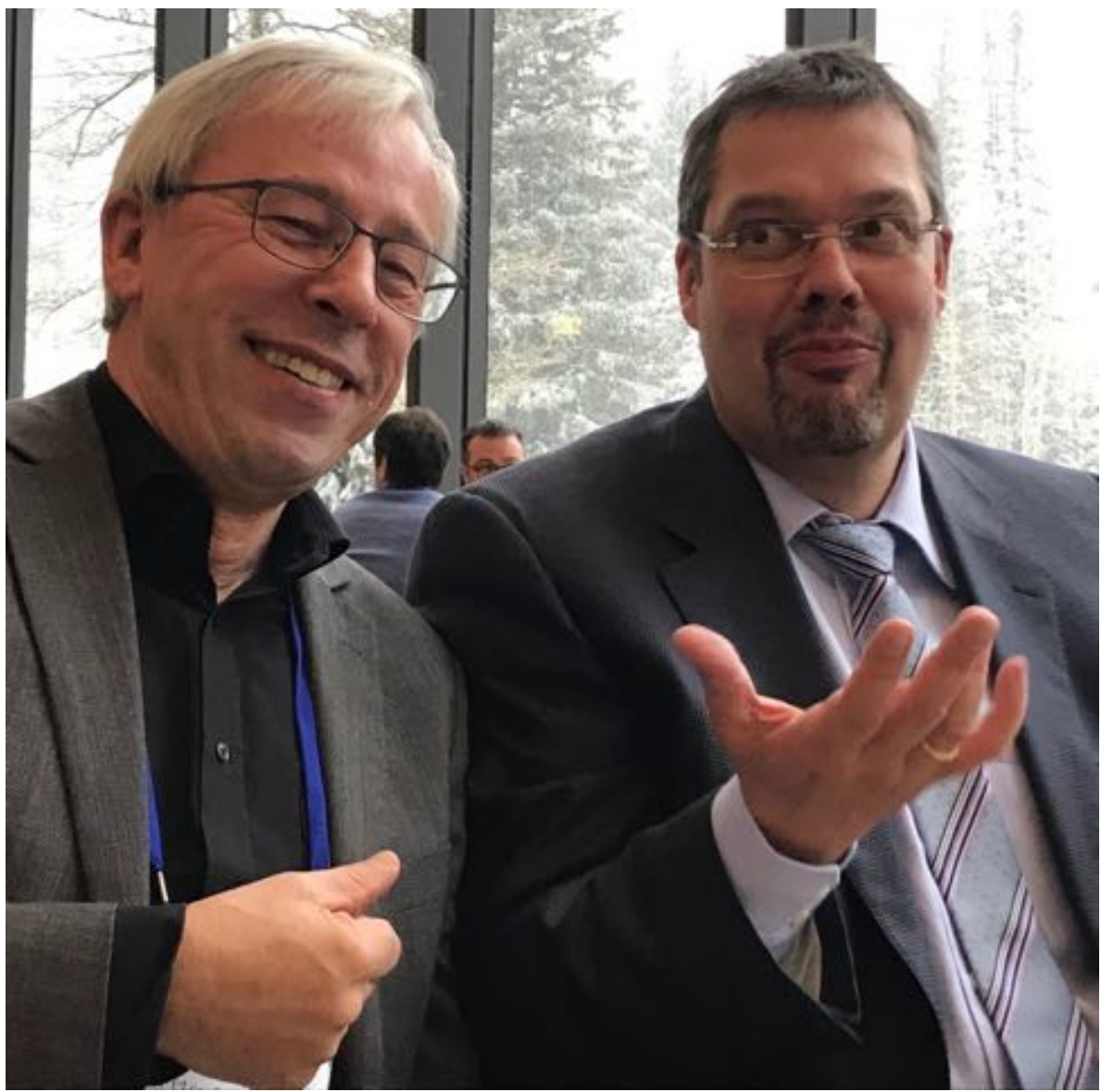

**Figure 15.** Ernst Rasel sharing a joke with Peter Zoller. Rasel and Zoller were co-recipients of the Willis E. Lamb Award. Zoller received the award for his groundbreaking work in the field of quantum optics and quantum information, whereas Rasel was honoured for his pioneering work in the field of ultra-cold atom research under absence of gravity. Credit: Suzy Lidstrom

that the moment has arrived where former speculations will start to be reality.

## 10. What does entropy mean and why is it so important? by Roland Allen

It is remarkable that entropy occurs nowhere in the most fundamental laws of physics, and yet is one of the most important concepts in all major areas of physics, as well as in other branches of science and technology. Some of the principal figures in understanding the need for a statistical description of macroscopic systems, and developing the idea of entropy, are shown in Figure 16, Figure 17 and Figure 18.

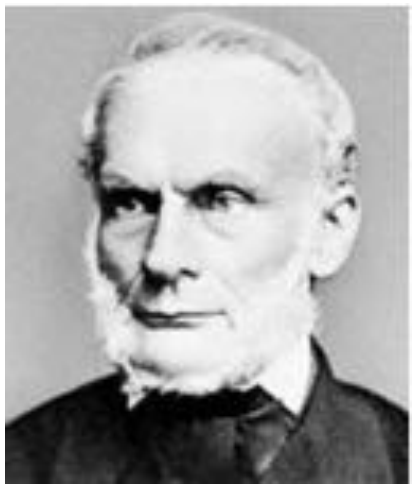

**Figure 16.** Rudolph Clausius introduced the concept of entropy. Credit: Wikimedia Commons.

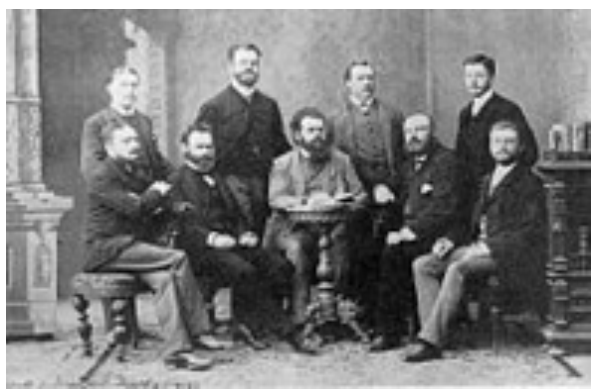

**Figure 17.** Standing, left to right: Walther Nernst, Heinrich Streintz, Svante Arrhenius, Richard Hiecke; sitting, left to right: Eduard Aulinger, Albert von Ettingshausen, Ludwig Boltzmann, Ignaz Klemencic, and Victor Hausmanninger, Graz, 1887. Source: Wikicommons; http://physik.kfunigraz.ac.at/itp/pictgal/pictgal.h

As Feynman points out in his book on statistical mechanics [126], the Gibbs entropy

$$S_G = -k \sum_i p_i \log\ p_i \tag{4}$$

never changes in a microscopic quantum description. Here $p_i$ is the probability that a system is in state $i$, and $k$ is the Boltzmann constant. This at first appears paradoxical, since the entropy usually increases in a realistic macroscopic description, in accordance with the second law of thermodynamics. The Gibbs entropy is the negative of the Shannon information, and is called the von Neumann entropy, after John von Neumann Figure 19, when it is re-expressed using the density matrix $\widehat{\rho}$, which is defined in terms of the states $|\Psi_j(t)\rangle$ with weights $w_j$:

$$S_G = -k\ tr\ \widehat{\rho} \log\ \widehat{\rho} \tag{5}$$

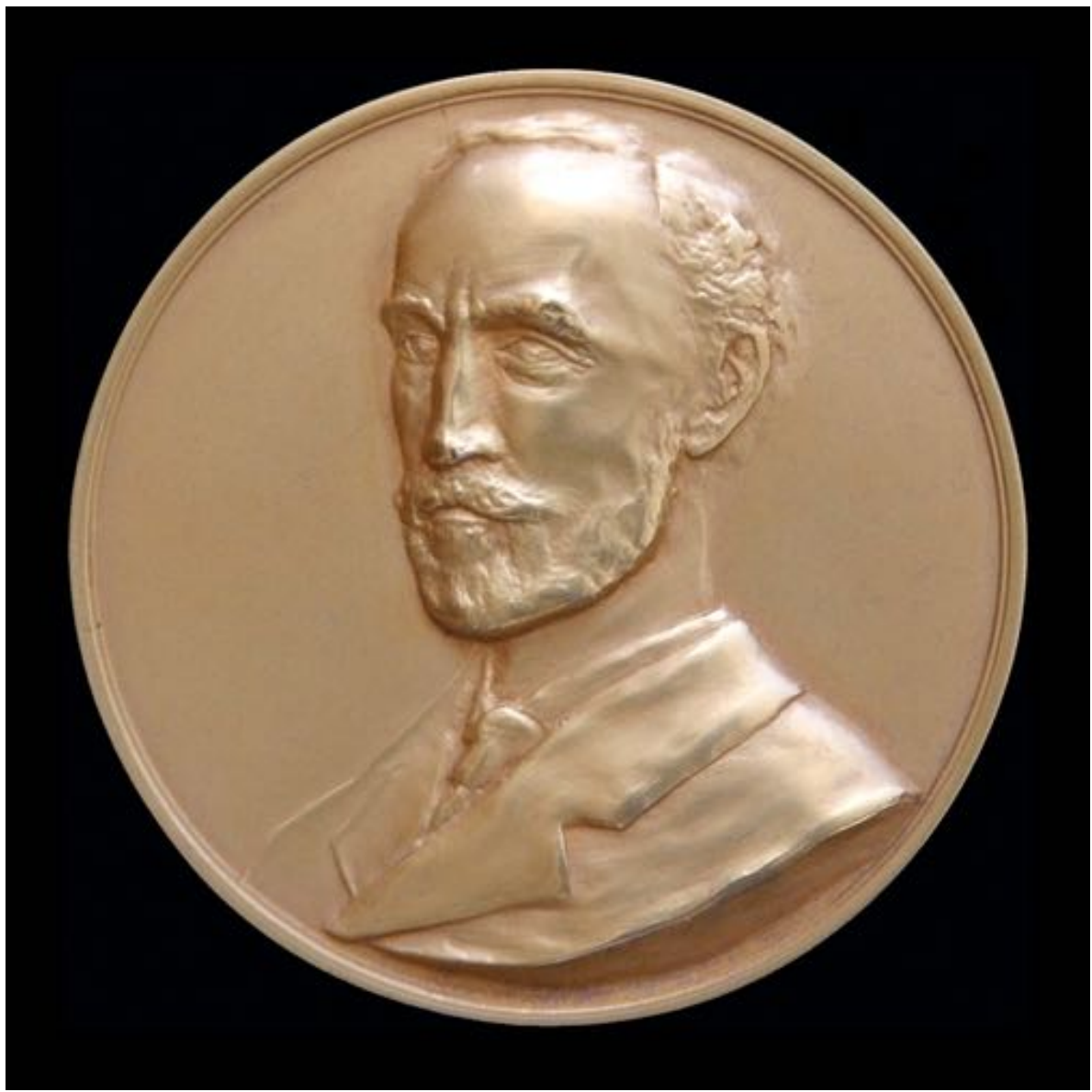

**Figure 18.** The Gibbs award, in honour of Josiah Willard Gibbs (1839–1903), is made: "To publicly recognize eminent chemists who, through years of application and devotion, have brought to the world developments that enable everyone to live more comfortably and to understand this world better." In 1911, Svante Arrhenius (Figure 18) received the first Gibbs award. The medal featured here was awarded to Linus Pauling, who remains the only person to have been awarded two undivided Nobel prizes (in chemistry and peace). Credit: Courtesy Ava Helen and Linus Pauling Papers, Oregon State University Libraries.

$\widehat{\rho} = \sum_j w_j \, |\Psi_j(t)\rangle \, \langle\Psi_j(t)|.$

The original statistical entropy of Boltzmann Figure 17

$$S_B = k \log W \tag{6}$$

similarly does not change in a microscopic description, but does change when the system moves to a macrostate with a larger number $W$ of microstates. The Gibbs entropy for an individual system can be derived from the Boltzmann entropy for an ensemble of systems in thermal equilibrium.

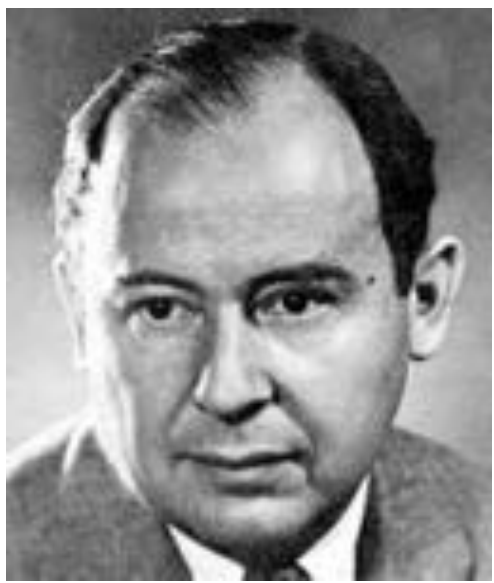

**Figure 19.** In addition to his work involving the density matrix and entropy, as well as fundamental aspects of quantum theory, John von Neumann made many groundbreaking contributions to pure and applied mathematics, the birth of modern computers, and widely varied areas of physics, including those required for success of the Manhattan Project. Edward Teller remarked that he was "one of those rare mathematicians who could descend to the level of the physicist." Source: Wikimedia Commons.

In both cases, the key point is that the entropy is a measure of our ignorance, when we can only observe and control the macrostates. A paradigm is a block sliding across a surface with friction. In either a classical or a quantum picture, a full microscopic description implies no loss of information or gain in entropy $= -$ information. In a macroscopic description, on the other hand, we retain only the information available with macroscopic variables. In order to have a quantitative treatment, we are then forced to introduce a new macroscopic variable – the entropy $S$, originally introduced by Clausius Figure 16 :

$$S_2 - S_1 = \int_1^2 \frac{dq}{T}. \tag{7}$$

This change in entropy is defined for a quasistatic process, which is envisioned as occurring so slowly that the system is always in thermal equilibrium as it moves between the macroscopic states 1 and 2. The increase $dE$ in the internal energy of the system is divided into the "work done by the system" $dw$ (the part that can be specified by macroscopic variables) and the "heat added to the system" $dq = TdS$ (the part that escapes description by macroscopic variables). Since $S$ is determined by the specified macroscopic variables, it also is a proper macroscopic variable.

For a non-quasistatic process – e.g., a sudden free expansion from state 1 to 2 – we can ordinarily still obtain $S_2 - S_1$ via a thought experiment which connects 1 to 2 through a different process which is quasistatic, since the entropy is determined by the macroscopic state.

Entropy has long been indispensable in fields ranging from industrial engineering to the exotic phases of materials, and it is now involved in truly remarkable mysteries at the current frontiers of physics and astrophysics, such as the Bekenstein-Hawking entropy of black holes [127] and the controversial related issue of whether information is lost within a black hole [128], [129].

The well-known attempts to explain black hole entropy have involved extremely sophisticated arguments in string theory, loop quantum gravity, etc. – and yet have still failed to explain the entropy of real black holes, which are not extremal and which dwell in three-dimensional space. These efforts are in dramatic contrast to the well-known simplicity of the formulae for the Bekenstein-Hawking entropy $S_{BH}$ and Hawking temperature $T_H$

$$S_{BH} = \frac{1}{4}\frac{A}{\ell_P^2}, \; T_H = \frac{1}{2\pi}\kappa \tag{8}$$

and the first law of black hole thermodynamics

$$dE = dq - dw, \; dq = T_H dS_{BH}, \; -dw = \Omega dJ \; + \; \Phi dQ \; . \tag{9}$$

Here $A$ is the surface area and $\kappa$ the surface gravity (at the event horizon); $\ell_P$ , $\Omega$, $J$, $\Phi$ and $Q$ are respectively the Planck length, angular velocity, angular momentum, electric potential, and charge.

Less often emphasized is the brilliant demonstration of Gibbons and Hawking [130]) that the Euclidean path integral $Z_{BH}$ of a general black hole yields exactly the right form for the Bekenstein-Hawking entropy, if it can be interpreted as a true thermodynamic partition function (ultimately based on microstates). It has recently been pointed out [131] that this is the case in a theory [132], with other merits including the prediction of a credible dark matter particle [133], [134], [135], in which the Euclidean action $S_E$ of any stationary system (including rotating black holes) fundamentally originates as an entropy $S$ . In this theory, Gibbons and Hawking (see Figure 20 and Figure 21) have explained the Bekenstein-Hawking entropy of all black holes – beginning with real black holes of any kind in a universe with three spatial dimensions, and with an argument that can even apparently be extended to higher dimensions.

There are many questions remaining: Were there primordial black holes created in the Big Bang which are hot enough to test the ideas of Hawking and others experimentally, perhaps even through observable explosions? Are the ideas of string theory or [132] valid? Are black holes in higher dimensions relevant?

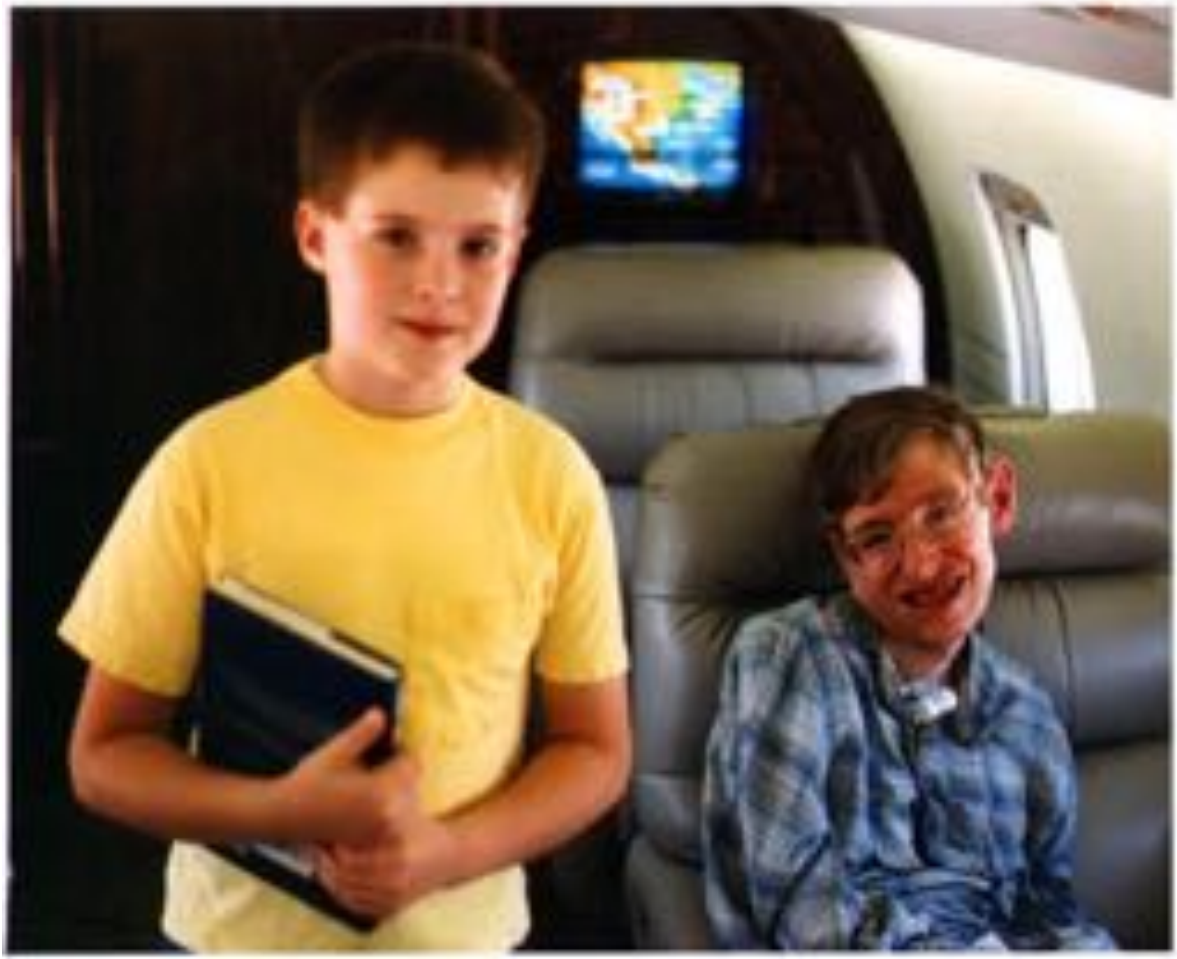

**Figure 20.** Stephen Hawking with a 10-year-old fan in 1995. The book is Hawking's *A Brief History of Time*, autographed with the author's thumbprint. The plane is an EDS corporate jet based in Dallas, which had flown in from Cambridge, England. Credit: Roland Allen.

In regard to the question of black hole information loss, two central issues are (i) the interpretation of quantum mechanics and (ii) the nature of quantum gravity, for which there is still no widely accepted theory. In [131] it is pointed out there is no loss of information provided that (i) one adopts the Everett interpretation of quantum mechanics, in which there is no magical collapse of the state of a system during a quantum measurement, and (ii) assumes that a quantum description of gravity will still describe the time evolution of states through a deterministic equation of the usual form

$$i\hbar \frac{\partial}{\partial t} |\Psi\rangle = \widehat{H} |\Psi\rangle \tag{10}$$

where $|\Psi\rangle$ specifies the state of all fields, including gravity, and $\widehat{H}$ is the operator that specifies the change in

$|\Psi\rangle$ during a time $dt$. There are clearly many questions remaining which can only be answered when there is a verified theory which describes all the quantum fields of nature, including gravity.

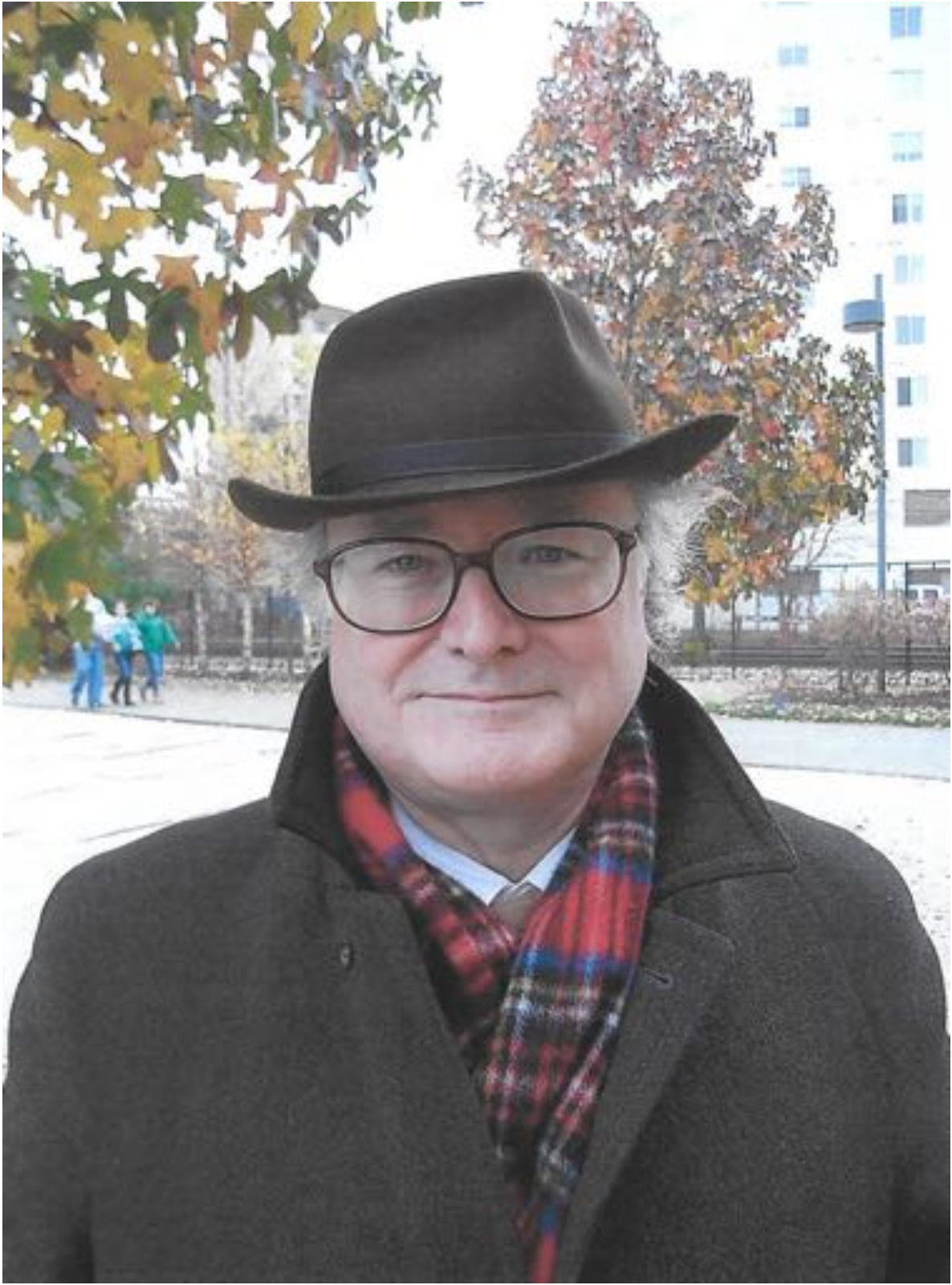

**Figure 21.** Gary Gibbons was a student of Stephen Hawking, obtaining his PhD from the University of Cambridge in 1973. He is a Fellow of the Royal Society, whose nomination notes that he "played a leading role in the development of the Euclidean approach to quantum gravity and showed how it could be used to understand the thermal character of black holes and inflating universes. This revealed a deep and unexpected relationship between gravitation and thermodynamics." Credit: Christine Gibbons.

It should be mentioned that there are many proposed extensions of entropy, motivated by plasmas and other astrophysical phenomena [136], nonequilibrium phenomena in general, mesoscopic systems, and other contexts where the standard ideas of thermodynamics and statistical mechanics are not directly applicable. In this short contribution it is impossible to discuss this large literature and related set of other unanswered questions. Perhaps the grandest aspiration is a generalisation of the existing principles of thermodynamics and statistical mechanics to biological systems, as in the next contribution by Mikhail Katnelson and Eugene Koonin.

## 11. Are there macroscopic variables that can usefully describe biological evolution? by Mikhail Katsnelson and Eugene Koonin

> A theory is the more impressive the greater the simplicity of its premises, the more different kinds of things it relates, and the more extended is its area of applicability. Therefore, the deep impression which classical thermodynamics made upon me. It is the only physical theory of universal content concerning which I am convinced that within the framework of the applicability of its basic concepts, it will never be overthrown.
>
> *Albert Einstein [137]*

This notable quote of Albert Einstein reflects the remarkable power of macroscopic, phenomenological approaches in physics. Indeed, "everything" in the world, or more precisely, any large ensemble of microscopic entities, obeys the laws of thermodynamics, in either the classical, equilibrium or the non-equilibrium form. Obviously, "everything" includes biological entities at different levels, from a cell to the biosphere for which the laws of thermodynamics play the role of basic constraints on function and growth. However, a far more interesting and, indeed, fundamental question is, if deeper parallels between thermodynamics and biology exist, and whether macroscopic variables analogous to the key variables of thermodynamics (temperature, energy, entropy) can be identified and become important descriptors of the biological evolution. We submit that there is indeed an isomorphism between phenomenological thermodynamics and biological evolution that can provide insights into the evolutionary process.

According to Dobzhansky's famous dictum, "Nothing in biology makes sense except in the light of evolution" [138] , and, as amended by Lynch with good justification, "Nothing in evolution makes sense except in the light of population genetics." [139]

Population genetics deals with large ensembles of biological entities at different levels, namely, numerous alleles in a genome and numerous individuals in a population. Hence parallels with phenomenological thermodynamics beg to be drawn. Indeed, in a general form, this connection was already recognized by Ronald Fisher, arguably, the most prominent founder of population genetics [140]. Many years later, the correspondence between thermodynamic and evolutionary variables was made explicit by Sella and Hirsch [141] , and further elaborated by Barton and colleagues [142] , and by ourselves [143] (Table 1).

**Table 1.** The isomorphism between the key macroscopic variables inthermodynamics and evolutionary biology. Credit: Reproduced from [143]. CC BY 3.0.

| Thermodynamic variable | Corresponding variable in evolutionary biology | |
| --- | --- | --- |
| | Sella and Hirsh [141] | Katsnelson, Wolf, Koonin [143] |
| Inverse temperature, $b=1/T$ | Effective population size ($N_e$) | Effective population size ($N_e$) |
| Entropy (per particle) | Derived from the free fitness expression | Evolutionary information density, D(N) (see main text) |
| Free energy Hamiltonian | -log (fitness) | - |
| Thermodynamic potential | Derived from the Hamiltonian using Gibbs formula | Evolutionary innovation potential, dI (see main text) |

A straightforward equivalency exists between effective population size, the key parameter of population genetics that governs the evolutionary regimes, and inverse temperature ($N_e \sim 1/T$). The analogy is complete and transparent: an infinite population (obviously, an abstract concept, but one routinely used in population genetics) is equivalent to 0 K (the ground state of a physical system). At low $T$, evolution is effectively deterministic because, under strong selection, only one, globally optimal genotype configuration survives. At the other end of the spectrum, at high $T$ (small $N_e$), evolution is a stochastic process that is dominated by random genetic drift. This regime engenders multiple evolutionary trajectories some of which cross valleys on the fitness landscape. Thus, counterintuitive as this might seem, innovation and emergence of complexity occur, primarily, in small, ostensibly, comparatively unsuccessful populations. It appears natural to introduce a new variable that is analogous to thermodynamic potential and can be interpreted as an evolutionary innovation potential:

A straightforward equivalency exists between effective population size, the key parameter of population genetics that governs the evolutionary regimes, and inverse temperature ($N_e \sim 1/T$). The analogy is complete and transparent: an infinite population (obviously, an abstract concept, but one routinely used in population genetics) is equivalent to 0 K (ground state of a physical system). At low T, evolution is effectively deterministic because, under strong selection, only one, globally optimal genotype configuration survives. At the other end of the spectrum, at high $T$ (small $N_e$), evolution is a stochastic process that is dominated by random genetic drift. This regime engenders multiple evolutionary trajectories some of which cross valleys on the fitness landscape. Thus, counterintuitive as this might seem, innovation and emergence of complexity occur, primarily, in small, ostensibly, comparatively unsuccessful populations. It appears natural to introduce a new variable that is analogous to thermodynamic potential and can be interpreted as an evolutionary innovation potential:

$\mathrm{d}\,I = \mathrm{d}\,t \left(\frac{\mathrm{d}\,S}{\mathrm{d}\,t}\right)/N_e$

Here, S is the evolutionary entropy [140] , [144] :

$$S = \sum_{i=1}^{L} S_i \ = - \sum_{i=1}^{L} \sum_j f_{ij} \ \log \ f_{ij} \,, \tag{11}$$

which is defined as the total entropy of the alignment of $n$ sequences of length $L$; $S_i$ is the per site entropy and $f_{ij}$ are the frequencies of each of the 4 nucleotides ($j$ = A, T, G, C) or each of the 20 amino acids in site $i$. Equation (11) corresponds to the classic Shannon entropy when applied to an alignment of homologous sequences rather than a single sequence (hence evolutionary entropy) [144] . Equation (11) does not take into account population parameters, in particular, $N_e$. However, as noted above, evolutionary innovations take place, primarily, in small populations where genetic drift results in a "free" movement of the population on the fitness landscape.

Equation (10) incorporates this pattern such that evolutionary innovation (rate of evolutionary entropy production) is inversely proportional to $N_e$. The calculations required to obtain the specific value of $dI$ can be complicated but, at least, in principle, these values can be derived by comparative genome analysis and correlated with other features, such as various measures of genomic complexity.

This view of evolution from the vantage point of phenomenological thermodynamics is in line with the theory of evolution of biological complexity that was developed by Lynch from population genetic considerations alone and shows that the genomic and organismal complexity of multicellular life forms evolves, primarily, via genetic drift in small populations [145]. Within the thermodynamic paradigm, the genomes of such organisms (in particular, animals and plants) are high temperature systems whose evolution is disorderly, and thus, leads to the emergence of complexity (for example, numerous genes interrupted by introns and producing multiple splice isoforms, large families of duplicated genes, and more). In contrast, the genomes of prokaryotes and viruses are orderly, low temperature systems in which complex, "baroque" features are weeded out by selection.

The thermodynamic parallels can be further extended to the model of major transitions in evolution (MTE) [146], [147]. The MTE include the origin of cells from pre-cellular life forms, origin of eukaryotes as a consequence of mitochondrial endosymbiosis, several cases of the origin of multicellular life forms, as well as origin of eusociality in animals and of superorganisms in plants and fungi. Each MTE involves the emergence of a new level and units of selection from ensembles of selection units at the preceding level (e.g. multicellular organisms evolving from collectives of cells. The isomorphism between phenomenological thermodynamics and population genetics translates into an interpretation of MTE as first-order phase transitions Figure 22 [143] . Typically, in such phase transitions, temperature remains constant but there is an entropy increase associated with the latent heat of transformation. As discussed above, in biological evolution, temperature corresponds to $1/N_e$ (Table 1) and, obviously, changes during MTE which evolve to increased size of evolutionary individuals. Effective population size inversely scales with the organism size, so the MTE are accompanied by an abrupt drop in $N_e$ or, in thermodynamic terms, rise in the evolutionary temperature. Instead, the quantity that remains nearly constant through the course of, and including the MTE, is the *evolutionary information density*:

$$D(N) = 1 - S/N \tag{12}$$

where S is evolutionary entropy (Equation (11) ), and L is the total number of sites in a genome.

In contrast, the evolutionary innovation potential increases during MTE, driven by the rising increased evolutionary temperature Figure 22 .

The equivalency between the macroscopic thermodynamic variables and the key variables of the evolutionary processes appears to reflect a deep commonality of the laws that govern the behavior of large ensembles of diverse entities. It is our belief that the thermodynamic perspective simplifies and clarifies the existing understanding of the evolutionary process, and hence, is at least conceptually useful. How much genuine advance in our understanding of evolution this approach can yield, remains to be investigated.

## 12. Can we observe the torsion of the connection in the geometry of the universe? by Philip Yasskin

When Einstein first introduced General Relativity, [148], [149] he assumed that the geometry was entirely determined by the metric. As a consequence, one is able to show that the geodesics determined by parallel transport are identical to those determined by minimizing distance. (The shortest distance between two points is a straight line.) Cartan [150], suggested that these two notions of geodesic might not be equivalent, that the covariant derivative (or connection) which determines the parallel transport might not be totally dependent on the metric which determines distances and angles. The part of the connection which is independent of the metric is called the torsion. So Cartan generalized Einstein's Theory to what is now called the Einstein-Cartan Theory [151], [152], [153] in which the metric and connection are independent, subject to the constraint that they are compatible, i.e. that the covariant derivative of the metric is zero.

Independently, Noether proved what is now called Noether's Theorem [154] which says that, in a Lagrangian based theory, each symmetry has a corresponding conservation law: Time and space translation invariance correspond to conservation of energy and momentum; rotation invariance corresponds to conservation of angular momentum; phase invariance in wave functions corresponds to conservation of electric charge; rotation invariance in isotopic spin space ( SU(2) ) corresponds to conservation of isotopic spin; and gauge invariance in the gauge group ( SU(3) ) corresponds to conservation of strong colour. In addition, each conserved quantity acts as the source for the force field which affects particles which possess that conserved quantity: Electric charge is the source for the electromagnetic field (photon, A) in the Maxwell equations and the electromagnetic field affects the motion of charged particles through the Lorentz force. Isotopic spin and color are the sources for the weak (W, Z) and strong (gluons) forces in the Yang-Mills equations and the weak and strong forces affect elementary particles through a generalization of the Lorentz force. Similarly, energy and momentum are the source for the metric in the Einstein equations and the metric affects the motion of all particles through the geodesic equation.

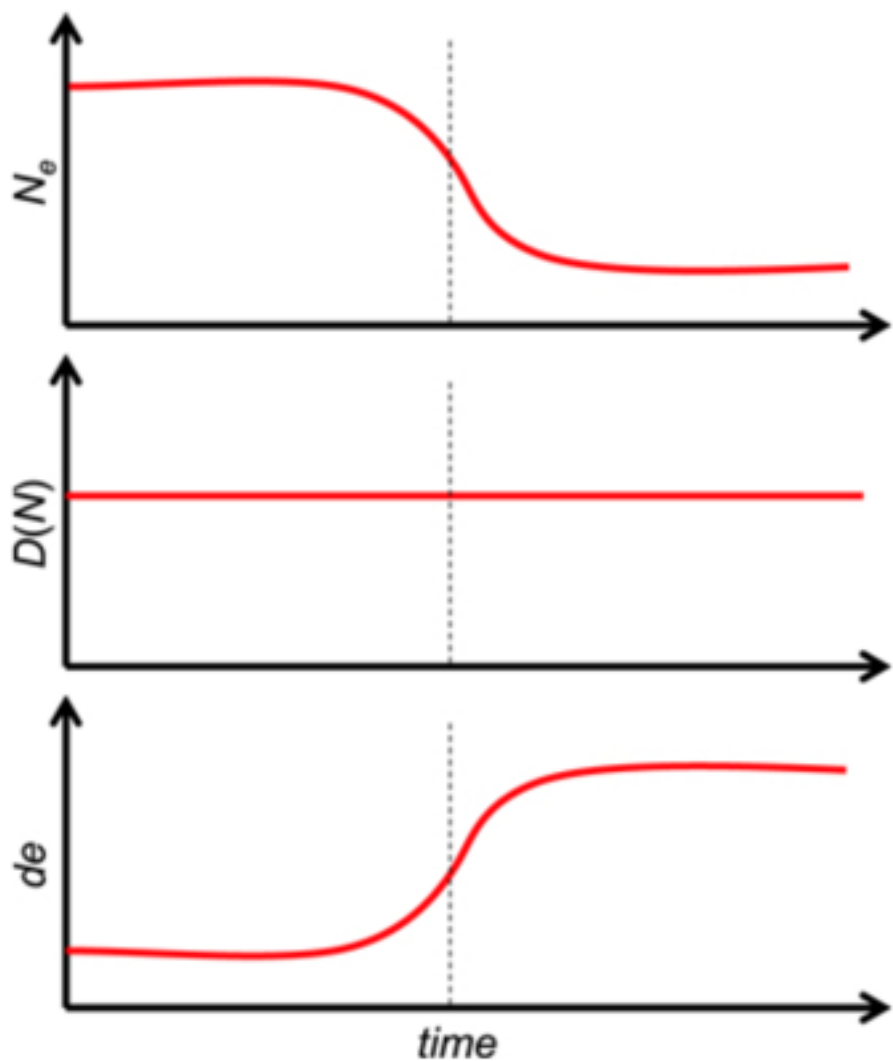

**Figure 22.** Major evolutionary transitions (MTE) as adiabatic first-order phase transitions. In each of the three panels, a MTE is denoted by a vertical dotted line, and the red curves show, conceptually, the change of the key parameters of the evolutionary process at MTE. Top panel: effective population size drops which corresponds to temperature rise; middle panel: evolutionary information density remains (approximately) constant; bottom panel: evolutionary innovation potential increases. Reproduced from [143]. CC BY 3.0.

It should be noticed that in Einstein's original theory, the conserved quantity, angular momentum, is not the source for anything, although it is still affected by the metric through the precession of gyroscopes. However, in Einstein-Cartan type theories, the angular momentum is the source for the connection in what should be called the Cartan equation and the connection affects the motion of particles through the parallel transport equation.

One indication of the acceptance of an independent connection, is that supergravity is based on the Einstein-Cartan theory, not just the Einstein theory, in that it allows for an independent connection (see van Nieuwenhuizen pages 197 and 205, [155]). So any attempts to unify gravity with the other field theories based on supergravity would probably do the same. The connection is the geometrical quantity that is most analogous to the potentials in other field theories. They both define the covariant derivative in the corresponding fiber bundle and they are the quantities one varies in the action to derive the field equations.

Since angular momentum is the source for the connection, one might expect that one could measure the torsion in the connection by observing the precession of a gyroscope such as that aboard the Gravity Probe B satellite. [156] Unfortunately, this is not the case. [157], [158], [159] A careful examination of the variational principle derivation of the field equations for the connection, shows that the source is actually just the spin angular momentum of elementary particles, not the orbital angular momentum of macroscopic bodies, perhaps because the latter is non-local. In particular, the angular momentum of a gyroscope (as on Gravity Probe B) is orbital. So its angular momentum cannot be a source for the torsion nor is its precession affected by the torsion.

So how can we measure the torsion of the connection? We can measure the precession of elementary particle spins. We can put an iron sphere in orbit, with uniform magnetization and watch the precession of the magnetization. In fact, it can be a null experiment. We can compare the precession of the magnet with that of the Gravity Probe B gyroscope. If there is any difference in their precessions, then that of the magnet cannot just be due to the metric; there must be torsion in the connection.

Suppose we want to perform this null experiment. The obvious thing to do would be to use a magnet whose total spin angular momentum is equal to the total orbital angular momentum of the Gravity Probe B gyroscope. Assuming the magnet is an iron sphere which is uniformly magnetized with one electron per atom aligned, then a back of the envelope computation shows that the sphere would need to be about ten meters in radius. Clearly, it is impossible to launch such a sphere into orbit. But some day, we may be able to mine the asteroids for iron and grow a single crystal in orbit with uniform magnetization. Further, our wonderful experimentalists may be able to find a way to use a smaller magnet and still be able to do a null experiment with the Gravity Probe B gyroscope. So it is not beyond imagination that in the next 100 to 200 years, we may be able to observe the torsion in the universe.

## 13. Can singularities in general relativity be resolved by quantum effects? by Alan Coley

**Singularity theorems and the cosmic censorship hypothesis:** Singularities occur within general relativity (GR), a geometric theory of gravity, both within black holes and at the beginning of the Universe at the big bang. The singularity theorems constitute one of the great theoretical achievements in classical GR (see the recent review [160]). Penrose's theorem [161] was the original singularity theorem, in which the important concept of geodesic incompleteness to characterize singularities [161] was introduced. Hawking subsequently realized that the conditions of this theorem are also satisfied in an expanding Universe to its past, and would then also lead to an initial singularity under reasonable conditions within GR, which led to the Hawking and Penrose singularity theorem [162] which formally states that:"If a convergence and a generic condition holds for causal vectors, and there are no closed timelike curves and there exists at least one of the following: a closed achronal imbedded hypersurface, a closed trapped surface, a point with re-converging light cone, then the spacetime has incomplete causal geodesics". However, since this theorem is proven using the strong energy condition, which might be violated even classically (e.g., by the matter fields present in the early Universe), the original Penrose theorem [161] that only utilizes the "null energy condition" is perhaps more relevant in the present context.

The singularity theorems imply the existence of spacetime singularities under rather general conditions, but they do not say very much about their properties. For example, although the well known Schwarzschild spherically symmetric vacuum black hole spacetime contains a singularity, it is shielded inside the so-called black hole event horizon and is therefore not visible to outside observers. This then led to the question [163], [164] of whether the gravitational collapse of physically realistic matter produces a singularity similar to that of Schwarzschild [161], in that it is concealed inside black hole event horizons (weak cosmic censorship) and is not timelike (strong cosmic censorship).

The weak cosmic censorship hypothesis states roughly that for Einstein's equations coupled to "physical" matter, no "naked singularity" will develop "generically" from nonsingular "realistic" initial conditions. Essentially, a naked singularity has the property that light rays can escape to infinity, so that the future is no longer theoretically predictable. In addition, it cannot be conjectured that naked singularities never occur, since there are known (albeit highly symmetric) examples. Since there can be no timelike singularities in a globally hyperbolic spacetime, a method for formulating (strong) cosmic censorship is as a statement that (under suitable conditions) spacetime must be globally hyperbolic. However, there are physically motivated spacetimes which are not globally hyperbolic (particularly in higher dimensions). The *cosmic censorship hypothesis is an important open question* (and was reviewed in [2]).

**What is the nature of cosmological singularities?:** Generic spacelike singularities are traditionally referred to as being cosmological singularities. Belinskii, Khalatnikov and Lifshitz (BKL) [165], [166] have conjectured that for a generic inhomogeneous cosmological model within GR, the approach to the (past) spacelike singularity is vacuum dominated, local and oscillatory, obeying the so-called BKL dynamics. In particular, due to the nonlinearity of the field equations of GR, if the matter is not a (massless) scalar field (e.g., it is a simple perfect fluid with a linear equation of state), then sufficiently close to the singularity *all matter terms can be neglected* in the field equations relative to dynamical geometrical terms (e.g., anisotropy). Numerical simulations have confirmed that the BKL conjecture is satisfied for special classes of spacetimes (see the review [167]). There have also been a number of theoretical approaches to study the structure of generic cosmological singularities, including the dynamical systems approach [168], [169]. Rigorous asymptotic mathematical proofs about general (Bianchi type IX and VIII) spatially homogeneous cosmological dynamics [170], [171], [172], [173], and the description of the generic asymptotic dynamics towards an inhomogeneous spacelike singularity in terms of an attractor [168], [169] have recently been presented.

Spike-like behaviour is a generic property of solutions of partial differential equations. Therefore, spikes are expected to occur in GR, and at exceptional points spatial derivatives do have an important effect (particularly, within cosmology, in the approach to the initial singularity in the oscillatory regime). As the cosmological singularity is approached, the spikes become narrower and narrower, and hence they are a significant challenge to study numerically [167]. Although some mathematical justification for spikes has been presented [174], [175], [176], actually obtaining an exact spike solution [177], [178] has been more successful. Numerical studies of so-called $G_2$ and more general cosmological models [179] have produced evidence that generally the BKL conjecture holds, except possibly on isolated surfaces where spikes form, and thus the asymptotic locality part of the BKL conjecture is violated.

**Is there a quantum theory of gravity?:** A fully consistent theory that subsumes both the incompatible theories of GR and the standard model of particle physics, which includes the quantum theories of electromagnetism and the strong and weak nuclear forces, is referred to as quantum gravity (QG). The question of whether string theory is a viable candidate for such a theory was reviewed in [180]. Although new physics result from QG modifications of GR, they do not appear to affect the macroscopic behavior of stellar systems and black holes very much. For example, perhaps the most important result of the unification of GR and quantum physics is the evaporation of a black hole via the emission of Hawking radiation [181]; but the behaviour of a classical blackhole does not change significantly over astrophysical timescales [182]. However, it is possible that QG may resolve the singularities of GR [2].

The very fact that singularities exist indicates that classical GR breaks down when the curvature of spacetime is sufficiently large. However, QG becomes important in such a regime, and hence is crucial to determine whether the singularity theorems are valid in the presence of quantum effects. The important question of whether solutions of GR can be extended beyond classical singular regimes within QG was first discussed in [183]. In any investigation of the singularity theorems within quantum gravity, it is important to first formulate the assumptions appropriately (such as, for example, averaged energy conditions relevant for the quantum region) and it is necessary to extend semi-classical theories to account for the quantum fluctuations of the spacetime itself.

In particular, since strings experience spacetime only through the so-called sigma model, it is plausible that spacetimes which are singular in GR can be regular in string theory. As noted earlier, in GR a singularity is defined in terms of geodesic incompleteness based on the motion of test particles; however, in string theory a spacetime is considered singular if test strings are not well behaved within the sigma model. A trivial example of a spacetime which is singular in GR but is not in string theory is the quotient of Euclidean space by a discrete subgroup of the rotation group. The resulting orbifold has a conical singularity at the origin, which leads to geodesic incompleteness in GR. However, it is completely regular in string theory since strings are extended objects. This orbifold has a very mild singularity, but even curvature singularities can be harmless in string theory. In addition, certain types of cosmological singularities can be smoothed out by closed string tachyon condensation [184].

However, it is not true that all singularities are removed in string theory. For example, string propagation in an exact plane wave string background can be studied and it can be shown that in some cases the string does not have well behaved propagation through the curvature singularity [185]. Moreover, a singularity in GR, such as the big bang or a black hole or naked singularity, is often characterised by the divergence of a physical or geometrical quantity as well as the breakdown of the evolution of geodesics. However, other types of singular behaviour can also occur due to pathologies of the tangent bundle (e.g., in conical singularities), or when there are directional singularities, in which the curvature diverges along some (but not necessarily all) directions. It is, of course, important to determine whether all singularities can be resolved within QG.

**Singularity resolution in GR by quantum effects:** Let us briefly review cosmological and black hole singularity resolution within loop quantum gravity (LQG) and string theory. LQG is a rigorous non-perturbative canonical quantization of gravity, in which the classical differential geometry of GR is replaced by a quantum geometry near the Planck scale. The application of LQG to cosmological (spatially homogenous) spacetimes is known as loop quantum cosmology (LQC), in which the infinite number of gravitational degrees of freedom reduce to a finite number. LQG suggests that generally singularities may be resolved by QG [186], [187]. In particular, due to the quantum geometry, the big bang is generically replaced by a "big bounce" which occurs without any violation of the energy conditions. A variety of spatially homogeneous cosmological models have been studied within LQC [188], [189] (see also [190]). The models that have been exactly solved have been shown to be well described by an effective theory that incorporates the main quantum corrections to the dynamics [191]. In particular, solutions of the effective equations for the generic (Bianchi type IX) spatially homogeneous spacetimes demonstrate numerically that the big bang singularity is resolved within LQC [192], [193].

It is possible that important features might be missed by greatly restricting the symmetry prior to quantization within LQC. However, it is expected that such investigations do in fact give valuable hints on loop quantization in inhomogeneous spacetimes [188], [189], [190]. Indeed, if the BKL conjecture is correct, it is anticipated that singularity resolution in simple spatially homogeneous models would at least capture some important aspects of the singularity resolution in more general (inhomogeneous) spacetimes. LQG techniques have been used to study the effects of QG in a class of very simple Gowdy inhomogeneous models [194]. However, the possible QG effects on spikes have not yet been studied.

Loop quantization of black hole spacetimes uses similar techniques to those of LQC, and leads to similar results on singularity resolution [195], [196]. The resolution of gravitational black hole singularities has also recently been studied within string theory [197], [198], [199]. In addition, some spacetimes exist in which singularities have been resolved by higher derivative corrections to the action [200], [201], [202]. For example, singularities were resolved in string solutions of five dimensional supergravity which include higher derivative supersymmetric corrections involving anomalies [203]. Such techniques to resolve singularities can applied in more general higher dimensional situations.

**Is there a quantum singularity theorem?** Gauge/gravity duality provides an alternative formulation of string theory in which asymptotically anti-de Sitter (AdS) boundary conditions are conjectured to be equivalent to a nongravitational quantum field theory (QFT) defined on the conformal boundary [204] (that is, the string theory lives in the 'bulk', while the QFT lives on the 'boundary'). An important consequence of holographic gauge/gravity duality is that results that cannot be demonstrated in one sector are often easier to show in the dual counterpart. Singularities (both cosmological and black holes) in the quantum realm have been investigated, and gauge/gravity duality has been used to study the validity of cosmic censorship with asymptotically AdS initial data (since it is not expected that the nature of singularities strongly depends on the asymptotic structure). Utilizing the *no transmission principle*, in which two quantum field theories whose Hilbert spaces do not overlap cannot transmit a signal to one another, some highly nontrivial consequences for holographic QG were deduced in [205], including the existence of a quantum version of cosmic censorship, that generic singularities inside black holes cannot be resolved and that a large class of bounces through cosmological singularities are forbidden. Therefore, although some singularities can be removed, a quantum singularity theorem is plausible.

In the classical singularity theorems certain positivity conditions on the stress-energy tensor such as the null energy condition are assumed which can beviolated locally in quantum field theory [206]. It is thus possible that in the highly quantum region near a big bang or black hole singularity, any possible negative energy might lead to the avoidance of the singularity. Therefore, it is of interest to ask whether there is a quantum mechanical generalization of any of the singularity theorems, which would make singularities inevitable even in quantum situations.

In [207] the so-called generalized second law (GSL) of horizon thermodynamics was proposed as a substitute for the null energy conditions in the standard singularity theorems of classical GR. Such a GSL is widely believed to hold as a consequence of the statistical mechanical properties of quantum gravitational degrees of freedom [208] (and has been demonstrated semi-classically for super-renormalizable fields with some extensions to general fields [209]. Therefore, the GSL is a reasonable candidate for a physical law likely to be valid in the quantum realm of a full theory of QG, thereby generalizing the notion of a trapped surface to a quantum deformed trapped surface in quantum situations. Indeed, it is known that the GSL implies a self-consistent semi-classical averaged null energy condition on achronal null geodesics in situations in which quantum effects are weak [210], [211]. In [207] it was shown that the (fine-grained) GSL can be used to prove the inevitability of singularities, extending the classical singularity theorem of Penrose [212] to the semi-classical setting. Therefore, not all singularities are resolved in QG. These conclusions are deduced in the context of semi-classical gravity, which is valid in regions of low curvature away from the singular region. However, in a region of high curvature, other than the GSL itself the results only rest on the fact that the basic notions of causality, predictability, and topological compactness continue to be meaningful in the theory of QG, and hence it has been argued that the conclusions will likely hold in a complete theory of QG [207].

## 14. How can a computer find autonomously new, surprising or creative solutions or insights? by Mario Krenn, Art I. Melvin and Anton Zeilinger

Automated computational methods have now been applied in situations within physics that might be considered rather unconventional, such as automated finding of laws of motion from experimental data [213], automated design of new quantum experiments [214], the application of machine learning in solving otherwise intractable quantum many-body problems [215], the identification of phases of matter [216], [217], and the discovery of new functional materials [218], [219] to name a few. All of these examples have a point in common: They search for solutions to a pre-defined question, a question defined by the human operator. Contemplating such computer-designed solutions can inspire new technologies or help discover surprising new connections [220], [221]. However, until today, a human has been needed to investigate the solutions in detail and decide whether they are *surprising* or *creative* - and whether one can learn something new from them.

The obvious question is:

**How can a computer find and identify autonomously new, surprising or creative solutions or insights?**

One requirement for a fact to be deemed surprising, or insightful, seems to be that it provides some new information. That has two consequences: 1) Surprisingness is subjective, 2) An algorithm needs to access substantial amounts of information about the domain where it should find surprising insights.

The algorithm could have large information provided by the humans in form of *big data*bases [222], [223], [224], [225]. At this point it seems very difficult to find a reasonable *surprisingness measure*. There is, however, exciting research being performed on this topic, such as the application of *Bayesian surprise* for specific domains [226] , [222] – but this has not been applied to scientific questions yet. One could even test whether the algorithm can correctly detect surprising information: Would a program with all the information that had been gathered by 1904 find that Einstein's photon hypothesis [227] was remarkable?

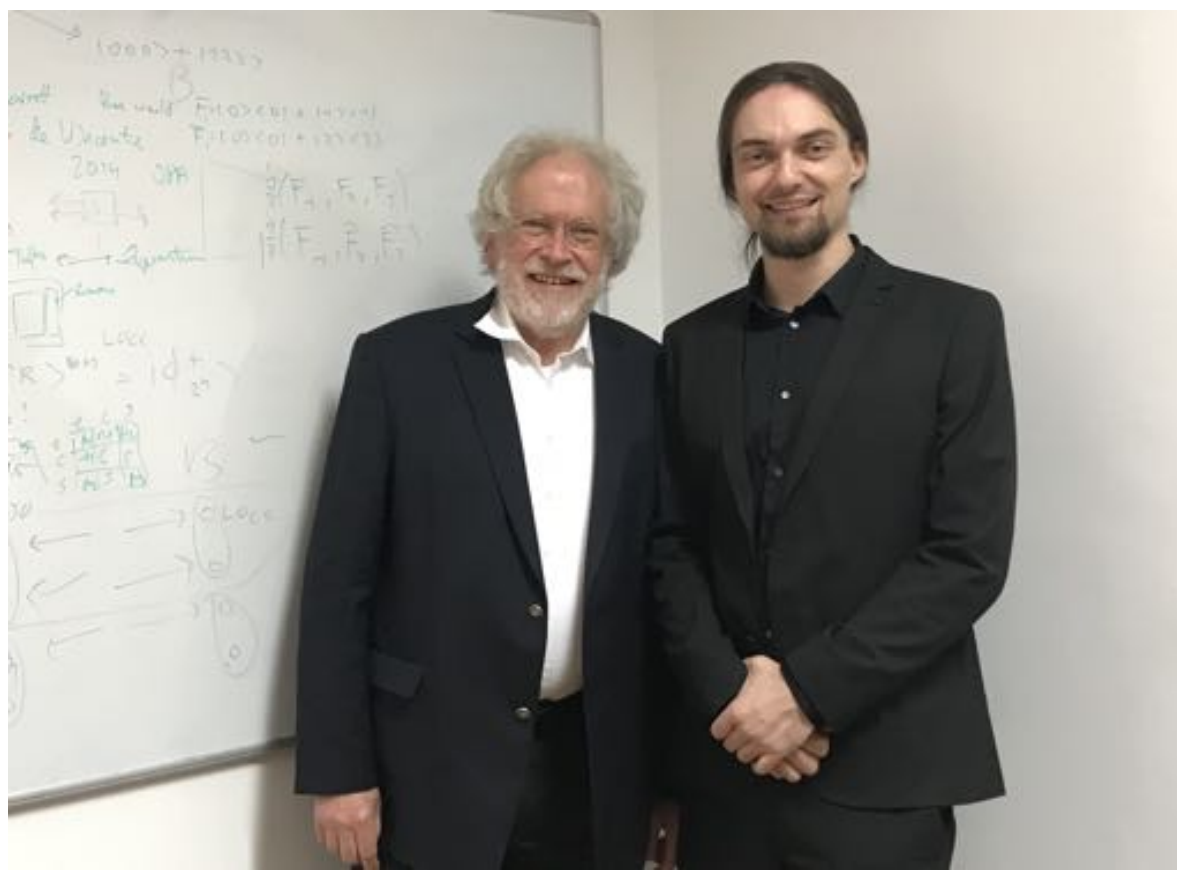

**Figure 23.** Anton Zeilinger, recipient of the sepcial medal of the senate, and Mario Krenn, on the occasion of the latter's thesis defense. Photographer: Xiaoqin Gao. Reproduced with permission.

Alternatively, the algorithm could try to gather the information about the domain itself. Famously, this has been applied in algorithms to play ancient Atari games [228] and and devised to learn playing the game of Go without any human knowledge [229]. This is possible in each of these examples as the game score can be provided during the game (for instance in Space Invaders) or the winner can be uniquely determined after the game (for instance in Chess or Go). With that, one finds out which moves are strategically successful. In science, such a simple external metric does not exist. A potential further step has been presented recently in the form of an algorithm that was able to start playing the game of Super Mario without receiving the game score. This was possible by applying an *intrinsic motivation* – the algorithm was building an internal model of its environment and chose actions which led to a maximal gain of information in the environment [230]. This method was called *curiosity-driven exploration*, and one could think that this program is an explorer or scientist in its little Super-Mario-world. Applying model-based algorithms (those which build a model of their environment) is technically very tricky and they are still the subject of basic research [231], [232], [233].

A quite different approach could be the application of entropy-based intrinsic motivations. In one impressive example, the algorithm tries to perform actions which maximizes its future possibilities. This information-theoretic idea applied to simple mechanical environments (such as the control of springs or balls) leads to the surprising emergence of quite complex behaviour [234]. This, in turn, leads to the question of whether interesting solutions or facts could also emerge from very general information-based criteria?

It would be amazing to find adequate methods which frequently lead to surprising new scientific insights or results. The hope is that automated application of such methods significantly speeds up our understanding of the world around us. Making progress in this question might requires a much better understanding of what human scientists are doing. However, perhaps searching for a *method* of doing science and attempting to reproduce it by means of a computer algorithm can not work [235] (*every methodology comes too late*) and something much more radical needs to be done before we are able to develop a computer algorithm which can explore our universe.

## 15. Guidelines for Including AIs as Co-Authors by Roman Yampolskiy

Recent progress in capabilities of Artificial Intelligence (AI) has raised a number of ethical questions [236]. One of the most interesting of these deals with the rights and responsibilities of AI as a contributor to scientific papers: "When will journals require that machines be listed on scientific collaborations?" Albeit somewhat tongue-in-cheek, one scientist has already voluntarily included AI as a co-author [237]. To respond to the question of author attribution, we need to answer two related questions: 1) Why do we formally list someone as a contributor to a paper? and 2) What are the capabilities an agent needs to have to be designated a contributor?

Co-author credit is given to acknowledge someone's effort fairly and to enable the person to receive any eventual recognition, fame, copyrights, promotion, future funding and other benefits attributable to the research, while at the same time assigning responsibility in the event that the work is found to be flawed or if questions arise. Consequently, for AI to be included as a contributing agent, rather than just acknowledged as a tool, it needs to be able to benefit from such inclusion and be capable of assuming moral responsibility for the work's shortcomings. Current AIs are known to be capable of performing only in narrow domains and do not yet have general intelligence as people do. In addition, today's AIs are not conscious; they experience neither pain nor pleasure, pride nor sadness, rendering them insentient to the traditional rewards of academic publishing, and also to the eventual penalties for a breach of trust.

Therefore, currently, it is not meaningful to include AI as a co-author. Nevertheless, it is important to acknowledge the use of AI because, in many cases, AI is implemented as a black box, arriving at correct decisions, but unable to explain how the answer was obtained. Failure to mention the contribution of AI may be misleading to the reader who would otherwise fail to understand how the results were obtained and verified. In the future it is predicted that we will develop Artificial General Intelligence (AGI) [238], a human comparable artificial intelligent system, and it has also been suggested that artificial consciousness will follow [239]. At that point, it would make sense to require including our artificial colleagues as official collaborators as they would be able to understand [the situation] and be morally responsible for any consequences of their efforts.

## 16. Interlude

We have confronted our audience with many instruments and voices in the first movement of this symphony, Sounds of Science. As it draws to a close, we consider the thoughts that Zdeněk Pa-poušek, chairman of the Committee on Education, Science, Culture, Human Rights and Petition of the Senate shared with us on the creative process Figure 24 in Prague in a short interlude.

> There are three ways of learning and three ways of expressing the truth: science, philosophy and art. The answer science gives is a full stop. The answer of philosophy is a question mark and the answer of art is an exclamation mark.

This, we were told, is because, philosophy "speculates and does not have to be ashamed if its answers end with a question mark", art should "appeal, alert and arouse, hence the exclamation mark", whereas "science does not like speculations, but prefers verified facts, hence the full stop". As scientists, however, we know not only that asking questions is vital when seeking answers, but that we are often only able to direct our research once we have determined how best to formulate the questions – in Roger Bacon's words, "To ask the proper question is half of knowing". And, indeed, like the questions that need framing, the answers might not come readily. As evidenced by the contributions presented here, scientists are sensible to the fact that science – like art and philosophy – is an ongoing endeavour. Thus, we suggest that, rather than a single fullstop, the science venture is, perhaps, best represented by an ellipsis: '...', and we conclude this first movement with the words: To be continued...

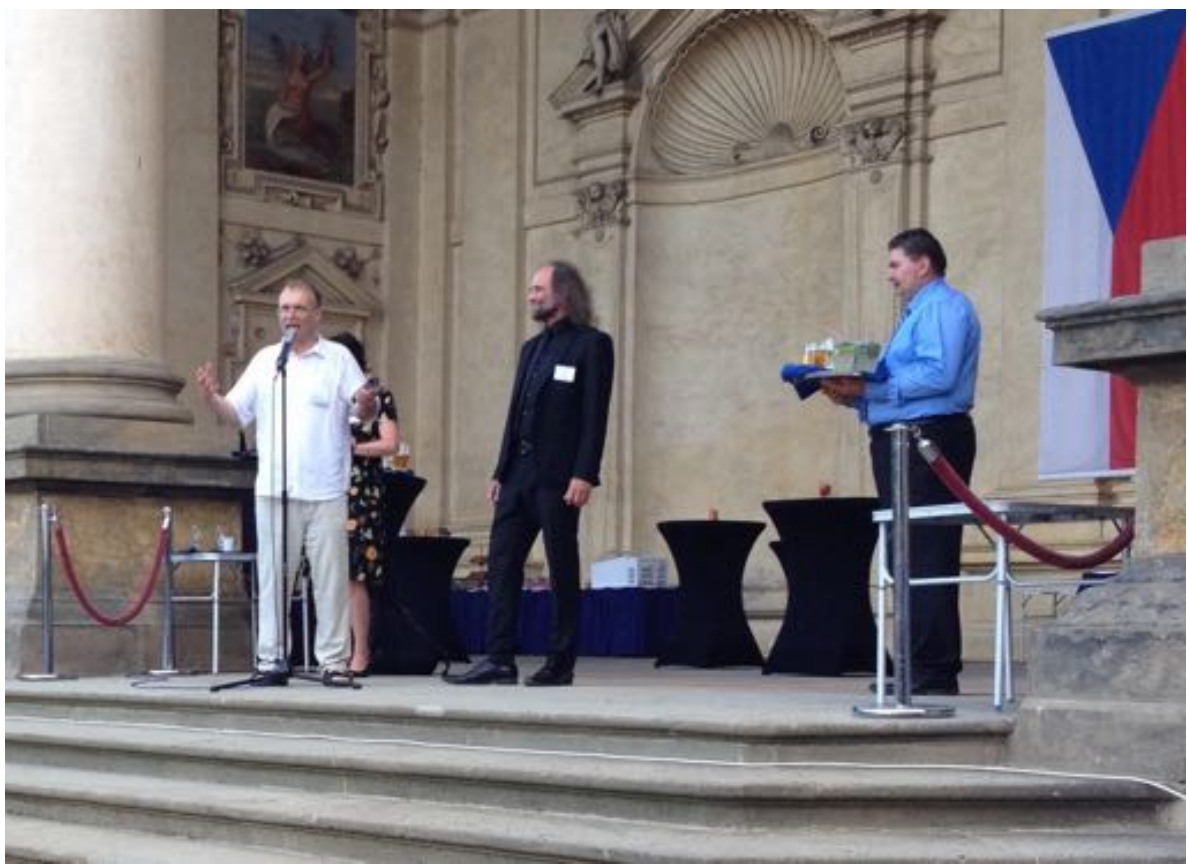

**Figure 24.** The conference organiser, Vaclav Spika, introducing Zdeněk Pa-poušek, the Chairman of the Committee on Education, Science, Culture, Human Rights and Petition of the Senate. Dr Pa-poušek informed the audience that there is a scientist, a philosopher and an artist within us all. He also said that: "Like the scientist, we need to test ideas and hold on to what is good; we should wonder and ask questions like a philospher, even without getting the answers; and we should perceive the world from the deepest corner of our soul and transform it into original artefacts in the spirit of an artist." Photograph: Suzy Lidström.

## Acknowledgements

Suhail Zubairy is thanked for proposing questions of such mind-blowing originality that we cannot begin to do them justice today, but which have, nevertheless, stimulated other – somewhat less challenging – thoughts outside the field of quantum optics, tackled here and in the second part of this paper. The following people are thanked for putting forward suggestions: Roland Allen, Dragos Anghel, Liliana Arachea, Howard B., Warwick Bowen, Ana Maria Cetto, Kuan-Hsun Chiang, Claudia Clarke, James Cresser, Bryan Dalton, Florian Fröwis, Ed Fry, Philippe Grangier, Johannes Hardstepne, Ivan Hartl, Peter Keefe, Stephan Klumpp, Mario Krenn, Theo Nieuwenhuisen, A. Pechen, David Sanchez, Hans Schussler, Marlan Scully, Tamar Seideman, Jurgen Stockburger, Behnam Tonekaboni, Vaclav Trojan, Joan V., Howard Wiseman, and Anton Zeilinger, as well as the researchers who proposed questions anonymously.

The authors are grateful to Gary Gibbons, Lucy Hawking, Robert Kirby of United Agents, and Oregon State University Libraries for assistance in obtaining the permission to use photographs. We would also like to thank everyone who appears in the photographs for agreeing to the use of their image in the context of an academic article.

S.L. would like to express her sincere gratitude to colleagues at Texas A&M University in College Station for many stimulating discussions, and the Department of Physics and Astronomy for hosting her during her sabbatical. She would like to acknowledge the enthusiasm with which this project has been greeted, and the input of many different researchers who posed questions for consideration in Prague, UAE and elsewhere.

The work presented by W.B. was supported by the Air Force Office of Scientific Research (grant number: FA2386-14-1-4046). P.G. would like to express his deep thanks to Alexia Auffèves and Nayla Farouki for many discussions and contributions. A.A.C. would like to thank Aron Wall for helpful comments and acknowledge the financial support provided by NSERC of Canada.

The staff of Typeset are thanked for technical assistance in producing the final draft of this manuscript.